\documentclass[letterpaper]{article} % DO NOT CHANGE THIS
\usepackage{aaai24}  % DO NOT CHANGE THIS
\usepackage{times}  % DO NOT CHANGE THIS
\usepackage{helvet}  % DO NOT CHANGE THIS
\usepackage{courier}  % DO NOT CHANGE THIS
\usepackage[hyphens]{url}  % DO NOT CHANGE THIS
\usepackage{graphicx} % DO NOT CHANGE THIS
\usepackage{natbib}  % DO NOT CHANGE THIS AND DO NOT ADD ANY OPTIONS TO IT
\usepackage{caption} % DO NOT CHANGE THIS AND DO NOT ADD ANY OPTIONS TO IT
\usepackage{algorithm}
\usepackage{algorithmic}
\usepackage{amsmath}
\usepackage{amsfonts}
\usepackage{amsthm}
\usepackage{subcaption}
\theoremstyle{definition}
\newtheorem{definition}{Definition}

\newtheorem{assumption}{Assumption}
\theoremstyle{remark}

\newtheorem{example}{Example}

\newcommand{\game}{\mathcal{G}} \newcommand{\agentset}{\mathcal{N}}
\newcommand{\edgeset}{\mathcal{E}} \newcommand{\actionset}[1]{S_{#1}}

\newcommand{\NE}{\mathbf{\bar{x}}}
\newcommand{\z}{\mathbf{z}} 

\newcommand{\zeros}{\mathbf{0}}

\newcommand{\x}{\mathbf{x}}

\newcommand{\epsX}{\epsilon_X}
\newcommand{\epsY}{\epsilon_Y}

\newcommand{\R}{\mathbb{R}}

\usepackage{newfloat}
\usepackage{listings}
\DeclareCaptionStyle{ruled}{labelfont=normalfont,labelsep=colon,strut=off} % DO NOT CHANGE THIS
\floatstyle{ruled}
\newfloat{listing}{tb}{lst}{}
\floatname{listing}{Listing}
\title{Stability of Multi-Agent Learning in Competitive Networks: \\ Delaying the Onset of Chaos}
\author {
    Aamal Hussain \textsuperscript{\rm 1},
    Francesco Belardinelli \textsuperscript{\rm 1}
}
\affiliations {
    \textsuperscript{\rm 1}Imperial College London \\
    aamal.hussain15@imperial.ac.uk, francesco.belardinelli@imperial.ac.uk
}

\usepackage{bibentry}
\begin{document}

\maketitle

\begin{abstract} 
    The behaviour of multi-agent learning in competitive network games is often studied within the context of 
    zero-sum games, in which convergence guarantees may be obtained. However, outside of this class the behaviour
    of learning is known to display complex behaviours and convergence cannot be always guaranteed. Nonetheless, in order
    to develop a complete picture of the behaviour of multi-agent learning in competitive settings, the zero-sum
    assumption must be lifted. 
    
    Motivated by this we study the Q-Learning dynamics, a popular model of exploration and exploitation in multi-agent learning, in competitive network games. We determine how
    the degree of competition, exploration rate and network connectivity impact the convergence of Q-Learning. To
    study generic competitive games, we parameterise network games in terms of correlations between agent payoffs and
    study the average behaviour of the Q-Learning dynamics across all games drawn from a choice of this parameter.
    This statistical approach establishes choices of parameters for which Q-Learning dynamics converge to a stable
    fixed point. Differently to previous works, we find that the stability of Q-Learning is explicitly dependent
    only on the network connectivity rather than the total number of agents. Our experiments validate these
    findings and show that, under certain network structures, the total number of agents can be increased without
    increasing the likelihood of unstable or chaotic behaviours. 
\end{abstract}

\section{Introduction}
Multi-Agent Learning in competitive games requires agents to maximise their individual, competing rewards whilst simultaneously exploring their actions to find optimal strategies. This leads to
a highly non-stationary problem where agents must react to the changing
behaviour of adversarial agents. The study of Multi-Agent Learning in competitive settings has achieved a number
of successes within the context of zero-sum games and their network variants
% \cite{hofbauer:zerosum,piliouras:cycles,piliouras:hamiltonian,ewerhart:fp,cai:minimax}.
These games model perfect competition between agents, yielding an underlying structure which make
them amenable for studying multi-agent learning. In particular it is known that certain learning dynamics asymptotically converge to an equilibrium in network zero-sum games \cite{ewerhart:fp,piliouras:zerosum,kadan:exponential} whilst others converge in time
average
\cite{anagnostides:last-iterate,mertikopoulos:finite,bailey:fast-furious}. 

Yet in practice the requirement of an arbitrary competitive game to exactly satisfy the zero-sum
condition is restrictive. It therefore becomes important to study the behaviour of learning agents
in arbitrary competitive games.The challenge in taking this step is that
there are infinitely many realisations of games which can be considered competitive, making a case
by case analysis intractable. Furthermore, the non-stationarity of learning in competitive games often leads to complex
behaviours such as cycles \cite{galla:cycles,piliouras:cycles} and even chaos
\cite{griffin:evonetworks,sato:rps}. In fact, recent work has shown that
chaotic dynamics occur in games even slightly perturbed from the zero-sum setting
\cite{galla:complex,sato:rps}. In addition, recent work \cite{hussain:aamas,sanders:chaos} has shown
that the ability of learning dynamics to reach an equilibrium with low exploration rates diminishes
as the number of agents increases. These technical challenges present a strong barrier towards ensuring the
convergence of learning in competitive games with many players.

However, in both \cite{hussain:aamas} and \cite{sanders:chaos} it was assumed that all agents are
directly influenced by all other agents in the environment. In practice, however, this does not hold.
Many ML applications, including Generative Adversarial Networks (GANs)
enforce structured interactions between models \cite{hoang:mgan,li:triple-gan}. Furthermore, real world problems such as robotic systems
\cite{hamann:swarm,Shokri2020Leader-FollowerActiveness}
and competitive game playing \cite{tuyls:stratego} impose a communication network
between agents. In economic settings, agents interact through social networks either online or in
communities.

\paragraph{Model and Contributions} Motivated by this, we study multi-agent
learning in \emph{network games}, in which interactions between agents are modelled by an underlying communication network. In this setting, we study the
\emph{Q-Learning} dynamic \cite{sato:qlearning,tuyls:qlearning}, a foundational model
for studying the behaviour of agents who explore their state space, whilst
simultaneously exploiting their rewards. 

To address the issue of studying generic competitive games, we take a
statistical approach towards our analysis which is inspired by the study of
ecological systems \cite{opper:phase,galla:random-replicator} and statistical
mechanics \cite{roudi:pathintegral,guili:mean-field}. Rather than
engaging in a case-by-case analysis, we parameterise competitive network games by the
strength of anti-correlation between agent payoffs. Then, we perform a kind
of \emph{average case} analysis over all games which are drawn from this
parameter. This process has shown a number of success in the analysis of
learning in games \cite{galla:complex,sanders:chaos,coolen:minority} and neural
networks \cite{coolen:stat-network,kadmon:chaos-network,crisanti:chaos-network}. 

Our analysis allows us to determine how the stability of the Q-Learning dynamics
is influenced by the competitiveness of the game and on the exploration rate. In
particular, we are able to define a \emph{stability boundary} in terms of these
parameters. We find that stable behaviours occur with low exploration rates in highly
competitive games, such as zero-sum games. However, as the game deviates further from perfect
competition, higher exploration rates are required to ensure convergent behaviours.
We also analyse how the network itself influences
Q-learning dynamics. We find that complex dynamics occurs frequently in strongly
connected networks as the number of agents increases. By contrast, there are
networks for which the total number of agents has no influence on the asymptotic
convergence of learning. 

The statistical approach requires taking the limit of large action spaces. As a result, our
theoretical stability boundary holds exactly in this limit. However, we evaluate its predictions
through rigorous numerical experiments in finite games, including representative examples from the
literature. We find that the experiments agree with the theoretical results and show that the
likelihood of complex learning dynamics depends explicitly on the network structure rather than the
total number of agents. In fact it is found that, as long as the network is chosen appropriately, an
arbitrarily large number of agents can be added to the multi-agent system without compromising
convergence of learning.

\paragraph{Related Work}

A number of recent advances in the theory of learning in games have drawn from
tools in evolutionary game theory
\cite{hofbauer:book,hofbauer:egd,tuyls:foundational-models}. Here, popular
learning algorithms such as Q-Learning \cite{sutton:barto},
Follow-the-Regularised-Leader \cite{shalev:online} and Fictitious Play
\cite{brown:fp} can be approximated by continuous time models
\cite{tuyls:qlearning,mertikopoulos:reinforcement}. Then, tools from
the study of continuous dynamical systems \cite{strogatz:book,meiss:book} can be
used to analyse the asymptotic behaviour of the learning dynamic. In this
manner, strong predictions can be made regarding convergence of learning in
games \cite{krichene:thesis,abe:MFTRL,bloembergen:review,Perolat2020FromRegularization}. Notable successes of
this method lie in \emph{network zero sum games}
\cite{cai:minimax,abernethy:hamiltonian} which models perfect competition
between agents. In this setting, it is known that a number of learning dynamics
converge asymptotically to an equilibrium \cite{piliouras:zerosum,ewerhart:fp,kadan:exponential}.

By contrast, few guarantees can be provided outside of this class
\cite{anagnostides:last-iterate}. In fact, complex behaviour such as limit
cycles \cite{imhof:cycles,galla:cycles,piliouras:cycles} and chaos
\cite{sanders:chaos,svs:chaos} are known to be prevalent in generic games. To
make progress on this front, learning in games has benefited from tools derived
from the study of \emph{disordered systems} \cite{roudi:pathintegral}. The premise
is that the exact choice of rewards in generic games has infinitely many possible
realisations. Therefore, it becomes necessary to parameterise the game and then
analyse the \emph{average} behaviour of the learning dynamic under all games
which share the same parameter. This analysis has been successful in the
analysis of ecological systems \cite{opper:phase,guili:mean-field}, Recurrent
Neural Networks, \cite{coolen:stat-network} and evolutionary game theory
\cite{coolen:minority,chowdhury:punishment}. 

Most similar to
our work are \cite{galla:complex,sanders:chaos}. In the former, the authors
analysed two player competitive games and studied Experience Weighted Attraction
(EWA) \cite{camerer:ewa}, a learning algorithm closely related to Q-Learning
\cite{piliouras:zerosum}.
They were able to derive a boundary in terms of game competitiveness and exploration rate
between stable learning dynamics and complex dynamics. In \cite{sanders:chaos},
the authors extended this work towards multi-player games in which each agent
interacts with all others. In this setting, it was shown the stability boundary
depends on the total number of agents in the game. In particular, the region in
which learning converges to fixed point seems to vanish as the number of agents
increases. This result is supported by that of \cite{hussain:aamas} in which a
lower bound on exploration rates was determined so that Q-Learning dynamics converge to a
unique equilibrium. Again it was shown that this lower bound increases with the
number of agents.

In this work, we refine the result of \cite{sanders:chaos} towards the setting
of generic \emph{network games}. Importantly, we find that the
stability boundary is independent of the total number of agents in the game, but
rather explicitly dependent on the connectivity of the network. 

\section{Preliminaries}

\subsection{Game Model} \label{sec::game-model}

A network polymatrix game (henceforth network game) is described by a tuple
$\game = (\agentset, \edgeset, \allowbreak (A^{kl}, A^{lk})_{(k, l) \in \edgeset})$. Here,
$\agentset$ denotes a set of agents indexed by $k = 1, \ldots, N$, and $\edgeset$
denotes the set of edges in an underlying network. In particular, $(k, l) \in
\edgeset$ if agents $k$ and $l$ are connected in the network. The set of
\emph{neighbours} of an agent $k$ is denoted by $\agentset_k = \{ l \in
\agentset \, : \, (k, l) \in \edgeset \}$. Associated with
each edge $(k, l) \in \edgeset$ are the payoff matrices $A^{kl}, A^{lk} \in M_n(\R)$,
where $n$ denotes the number of actions an agent can play. The set of actions playable by agent $k$ is indexed by $i = 1, \ldots, n$. The strategy $\x_k$ of agent $k$ is a probability distribution over their set of actions and so is
chosen from the $n$-simplex $\Delta_n = \{\x \in \R^n \, : \, \sum_i x_{ki} = 1,
x_{ki} \geq 0\}$. Then for any agent $k$, given the joint strategy $\x_{-k}$ of
their opponents, their total reward $r_{ki}$ is given by 

\begin{equation}
    r_{ki}(\x_{-k}) = \sum_{(k, l) \in \edgeset} \left( A^{kl} \x_l \right)_i
\end{equation}
In words, it is the sum of the payoff they receive in each of the games played
along their edges. With this in place we can define an equilibrium for the game
as follows.
\begin{definition}[Quantal Response Equilibrium (QRE)] A joint mixed strategy $\NE \in \Delta$ is a
\emph {Quantal Response Equilibrium} (QRE) if, for all agents $k$ and all actions $i \in
\actionset{k}$
    \begin{equation*}
        \NE_{ki} = \frac{\exp(r_{ki}(\NE_{-k})/T)}{\sum_{j \in \actionset{k}} \exp(r_{kj}(\NE_{-k})/T)}.
    \end{equation*}
    where $T \in [0, \infty)$ denotes the \emph{exploration rate} of all agents.
\end{definition}
The QRE \cite{camerer:bgt} is the prototypical extension of the Nash Equilibrium to the case of
agents with bounded rationality, parameterised by the \emph{exploration rate} $T$. In
particular, the limit $T \rightarrow 0$ corresponds exactly to the Nash Equilibrium, whereas
the limit $T \rightarrow \infty$ corresponds to a purely irrational case,
where the QRE is unique and lies at the uniform distribution \cite{mckelvey:qre}.

\paragraph{Payoff Correlations}
As mentioned in the introduction, the entries of $A^{kl},
A^{lk}$ can take any value in $\R$, making a case-by-case analysis intractable.
We therefore move towards a kind of \emph{average case} analysis. In particular,
we construct ensembles of games at random which are parameterised by the strength of anti-correlation between opponent payoffs. Then we can analyse the \emph{expected}
behaviour of Q-Learning dynamics for different choices of this parameter. This
approach has yielded a number of successes in analysing replicator dynamics
\cite{opper:phase} and learning in games \cite{galla:complex,sanders:chaos}.

Averaging over the infinite possibilities of payoff matrices which could arise in a game theoretic
setting has the immediate effect of reducing the information available regarding the effect of the
payoffs on stability. However, the primary concern of this work is to understand
the effect of the network structure on stability. As such, it makes sense to
average over other factors. Indeed, the relevance of the payoff matrices on
the learning dynamics is an open and important topic for research and we point the interested reader
to \cite{pangallo:bestreply,pangallo:taxonomy} for rigorous treatments on the matter.

We must then ask how best to parameterise the payoffs. We do this by invoking the \emph{maximum
entropy} principle which is foundational to statistical mechanics
\cite{galla:complex,roudi:pathintegral}. This states that the natural choice for the payoff matrices
is that which maximises entropy subject to given conditions. In particular, these conditions are
\begin{align}\label{eqn::payoffcorrelations}
    \mathbb{E}[A^{kl}_{ij}] &= 0, \hspace{0.5cm} \forall k \in \agentset, \forall i, j \in \actionset{k}  \nonumber\\
    \mathbb{E}[\left(A^{kl}_{ij}\right)^2] &= 1, \hspace{0.5cm} \forall k \in \agentset, \forall i, j \in \actionset
     {k} \\
    \mathbb{E}[\left(A^{kl}_{ij}\right) \left(A^{lk}_{ji}\right)] &= \Gamma, \hspace{0.5cm} \forall l \in N_k, \forall
     i \in \actionset{k}, j \in \actionset{l} \nonumber
\end{align}

Intuitively, these conditions enforce that payoffs have zero mean and positive variance and,
crucially, enforces a correlation between the payoffs between two connected agents $k, l$
parameterised by $\Gamma \in [-1, 0]$. In the Supplementary Material we discuss, as an example, games which are drawn with $\Gamma = -1$. Here, the payoffs to each agent are exactly negatively correlated, corresponding to a zero-sum game. By contrast, when $\Gamma = 0$, the payoffs are completely uncorrelated. As such, $\Gamma$ controls the
\emph{competitiveness} of the game. In line with the maximum entropy argument and previous literature
\cite{galla:complex,sanders:chaos}, we draw the payoff matrices from a multivariate Gaussian
distribution with mean and covariance defined as in (\ref {eqn::payoffcorrelations}). A careful
treatment on why the multivariate distribution satisfies the maximum entropy argument can be found
in \cite{galla:complex}. Furthermore, as we show in our experiments(and in line with
previous studies on the analysis of random games \cite{galla:complex,sanders:chaos}), the predictions made in
the average case analysis of random games carries over strongly in experiments. 

% \begin{example} As an example we consider a class of two-player two-action games, which corresponds to a network game with a
% single edge associated with the payoff matrices 
% \begin{equation*}
%     A = \begin{pmatrix}
%         a_{11} & a_{12} \\
%         a_{21} & a_{22}
%     \end{pmatrix}, \; 
%     B = \begin{pmatrix}
%         b_{11} & b_{21} \\
%         b_{12} & b_{22}
%     \end{pmatrix}
% \end{equation*}
% Let $A, B$ be drawn from the ensemble of games generated with $\Gamma = -1$, i.e. the payoffs are drawn from a
% multivariate gaussian $p \sim N(0, \Sigma)$ where
% \begin{align*}
%     p &= \begin{pmatrix}
%         a_{11} & a_{12} & a_{21} & a_{22} & b_{11} & b_{21} & b_{12} & b_{22}
%     \end{pmatrix}^\top,   \; \\
%     \Sigma &= \begin{pmatrix}
%         1 & 0 & 0 & 0 & -1 & 0 & 0 & 0 \\
%         0 & 1 & 0 & 0 & 0 & -1 & 0 & 0 \\
%         0 & 0 & 1 & 0 & 0 & 0 & -1 & 0 \\
%         0 & 0 & 0 & 1 & 0 & 0 & 0 & -1 \\
%         -1 & 0 & 0 & 0 & 1 & 0 & 0 & 0 \\
%         0 & -1 & 0 & 0 & 0 & 1 & 0 & 0 \\
%         0 & 0 & -1 & 0 & 0 & 0 & 1 & 0 \\
%         0 & 0 & 0 & -1 & 0 & 0 & 0 & 1 
%     \end{pmatrix}
% \end{align*}
% Then, $a_{11} = -b_{11}$, $a_{12} = -b_{21}$ and so on. This precisely defines a zero-sum game which models a
% purely competitive game. Thus, the choice $\Gamma = -1$ captures all pairwise zero-sum games. By contrast, when $\Gamma = 0$, the payoffs are completely uncorrelated. As such, $\Gamma$ controls the
% \emph{competitiveness} of the game being played.
% \end{example} 

\subsection{Learning Model} \label{sec::learning-model}

    In this work, we analyse the \emph{Q-Learning dynamic}, a prototypical model for determining
    optimal policies by balancing exploration and exploitation. In this model, each agent $k \in
    \agentset$ maintains a history of the past performance of each of their actions. This history is
    updated via the Q-update:
    \begin{equation*}
        Q_{ki}(\tau + 1) = (1 - \alpha_k) Q_{ki}(\tau) + \alpha_k r_{ki}(\x_{-k}(\tau)),
    \end{equation*}
    where $\tau$ denotes the current time step, and $Q_{ki}(\tau)$ denotes the \emph{Q-value} maintained
    by agent $k$ about the performance of action $i \in S_k$. In effect $Q_{ki}$ gives a discounted
    history of the rewards received when $i$ is played, with $1 - \alpha_k$ as the discount factor.
    
    Given these Q-values, each agent updates their mixed strategies according to the Boltzmann
    distribution, given by
    \begin{equation*}
        x_{ki}(\tau) = \frac{\exp(Q_{ki}(\tau)/T) }{\sum_j \exp(Q_{kj}(\tau)/T)},
    \end{equation*}
    in which $T \in [0, \infty)$ is the \emph{exploration rate} of all agents.
    
    It was shown in \cite{tuyls:qlearning,sato:qlearning} that a continuous time approximation of
    the Q-Learning algorithm could be written as
    \begin{equation} \tag{QLD} \label{eqn::QLD}
        \frac{\dot{x}_{k i}}{x_{k i}}=r_{k i}(\x_{-k})-\langle \mathbf{x}_k, r_k(\x_{-k}) \rangle +T \sum_{j \in S_k} x_{k j} \ln \frac{x_{k j}}{x_{k i}},
    \end{equation}
    which we call the \emph{Q-Learning dynamics} (QLD). The fixed points of this dynamic coincide
    with the QRE of the game \cite{piliouras:zerosum}. We can rewrite the dynamic in the following
    form
\begin{align}
    \frac{\dot{x}_{ki}}{x_{ki}}&=r_{ki}(\x_{-k}) - T \ln x_{kj} - \rho_k  \label{eqn::QLDynamics} \\
    \rho_k &= \langle \x_k, r_k(\x_{-k}) \rangle - T \langle \x_k, \ln \x_k \rangle \nonumber \\ 
    r_{ki}(\x_{-k}) &=  \sum_{l \in N_k} \left(A^{kl} \x_l \right)_i.
\end{align}
In this light, $\rho_k = \rho_k(\x(t))$ can be seen as a normalisation factor which ensures that
$\sum_{i \in S_k} x_ {ki} = 1$. In addition, it is clear that the behaviour of the Q-Learning
dynamics depend strongly on the choice of $T$, the structure of the edge set
$\edgeset$, and the payoffs $A^{kl}$. In \cite{sanders:chaos}, a variant of (\ref{eqn::QLD}) was analysed
in which the concept of the interaction network was not introduced. Rather, each agent would be
required to play a single $p$-player game against all other agents in the environment. In their
work, it was found that complex dynamics (cycles and chaos) becomes more prominent as the number of
players in the game increases. This is similarly true of \cite{hussain:aamas} in
which (\ref{eqn::QLD}) was analysed in arbitrary games, without imposing any
structure on interactions between agents. The authors similarly concluded that,
as the number of agents increases, higher exploration rates $T$ are required
to ensure convergence to a QRE.

The introduction of a communication network allows for the interactions between agents to be included in the study. In particular, we determine how the number of neighbours for each agent affects the stability of (\ref{eqn::QLD}) and
compare this with the total number of players. In this light, we make the following assumption.

\begin{assumption} \label{ass::regular-network}The agents all share the same number of neighbours $N_0$,
i.e., $N_0 = |\agentset_1|, \ldots, |\agentset_N|$. In graph theoretic terms, we require that the network is
\emph{regular} \cite{egerstedt:graph-ctrb}.
\end{assumption}   

Assumption \ref{ass::regular-network} allows us to parameterise the connectivity of the network using
only the number of neighbours $N_0$. Additionally it allows us to make a direct comparison of our
results with that of \cite{sanders:chaos} who study the case where all agents are connected, in our case 
this is $N_0 = N-1$. Whilst the study of regular networks is certainly well motivated in the literature on
multi-agent systems \cite{egerstedt:graph-ctrb,mesbahi:graph-ctrb,fax:consensus}, a fruitful direction for extending
the work in this paper would be to introduce graph theoretic parameterisations (e.g. norms on the adjacency matrix)
to include heterogeneously coupled networks in the analysis.

\section{Statistical Analysis of Learning in Networks} \label{sec::main-result}

In this section, we derive a necessary condition for (\ref{eqn::QLD}) to
converge in generic network games. The process for doing this is as follows.
First, we average over the ensemble of games drawn using a particular choice of
$\Gamma$. This allows us to define an \emph{effective dynamic}, which reflects
the expected behaviour of (\ref{eqn::QLD}) over all games within the ensemble.
Then, we determine a necessary condition for a fixed point of this dynamic to be
stable. Similar calculations have been used to analyse replicator
dynamics \cite{galla:random-replicator,opper:phase} and variants of Q-Learning
\cite{sanders:chaos,galla:complex}, as well as minority games
\cite{coolen:minority} and recurrent neural networks \cite{coolen:stat-network}. By
following this process, we derive an estimate for the boundary between stable
fixed points and other behaviours, such as limit cycles and chaos. This boundary
is determined with respect to the payoff correlations $\Gamma$ and the
exploration rate $T$.

The analytic result requires applying techniques from statistical mechanics, where the theory
holds exactly only in the limit of large payoff
matrices, i.e., the number of actions $n \rightarrow \infty$. As shown in this Section the method allows us  to isolate
precisely how the number of neighbours $N_0$ and the total number of players $N$
affects the stability of learning in arbitrary competitive games. The limitation
is that the analytic result will overestimate the stability boundary for finite
games. However, our numerical experiments show that the predictions made in the
limit hold in practice for finite games. 

\subsection{Effective Dynamics}

In our first step, we derive an \emph{effective dynamics} which describes the
expected behaviour of the Q-Learning dynamics averaged over all possible
assignments of payoff matrices. This calculation is lengthy, so we report the
full details in the Supplementary Material. The idea is to define a probability
measure (Generating Functional described in the Appendix) over trajectories generated by Q-Learning dynamics, given some payoff matrices.
Next, we invoke the assumption that the payoff matrices are drawn from a
multivariate Gaussian with correlations parameterised by $\Gamma$. Using this,
we find the average form of the probability measure over all games generated with a 
choice of $\Gamma$, which we call its \emph{effective} form.  It is here that we use the limit of large payoff matrices (number of actions $n \rightarrow \infty$). Finally, we identify the effective probability measure with an associated dynamical system, which we call the \emph{effective} form for the Q-Learning dynamics. Intuitively, the \emph{effective Q-Learning dynamics} (\ref{eqn::effectivedynamics}) describes the average trajectories of Q-Learning over all, possibly infinite, games which are generated from a choice of $\Gamma$.
{\small
\begin{equation} \label{eqn::effectivedynamics}
    \frac{\dot{x}(t)}{x(t)} = N_0 \Gamma \int dt' G(t, t') x(t') - T \ln x(t) - \rho(t) + \sqrt{N_0} \eta(t),
\end{equation} 
}
Here, $\eta(t)$ is a Gaussian random variable which satisfies $\langle \eta(t), \eta(t') \rangle_* = \langle x(t), x(t') \rangle_*$ and $\langle \eta(t) \rangle_* = 1$. Following \cite{opper:phase}, we use $\langle \cdot \rangle_*$ to denote an average taken over all possible realisations of payoffs drawn using a choice of $\Gamma$. Similarly, $G(t, t')$ is a random variable satisfying $G(t, t')=\left\langle\frac{\delta x(t)}{\delta \eta(t')}\right\rangle_*$  As such, $G$ and $\eta$ capture the time correlations between the strategy at times $t$ and $t'$. We assume, as part of our derivation, that the initial
conditions for all agent strategies are independently and identically distributed (i.i.d) and so we drop the distiction between agents $k$ and actions $i$ in (\ref{eqn::effectivedynamics}).

\subsection{Stability Analysis}

Next, we determine the stability of fixed points for the effective dynamics. To
do this, we write $x(t) = \bar{x} + \tilde{x}(t)$ where $\bar{x}$ denotes a fixed
point of (\ref{eqn::effectivedynamics}) and $\tilde{x}(t)$ denotes perturbations
due to an additive white noise term $\xi(t)$ which is drawn from a Gaussian
of zero mean and unit variance.  Similarly we write $\eta(t) = \bar{\eta} +
\tilde{\eta}(t)$ where $\tilde{\eta}$ gives perturbations in the time correlation variable due to the random noise. The problem of determining stability of fixed points is now a
question of the growth or decay of $\tilde{x}(t)$ close to the fixed
point $\bar{x}$. To do this, we only need to keep terms in (\ref{eqn::effectivedynamics}) which are linear in
$\tilde{x}(t), \tilde{\eta}(t), \xi(t)$. This yields
{\small
\begin{align}\label{eqn::perturbeddynamics}
    \frac{d}{dt}\tilde{x}(t) &= -T \tilde{x}(t) \\ &+ \bar{x} \left(N_0 \Gamma \int dt' G(t-t') \tilde{x}(t') + \sqrt{N_0} \tilde{\eta}(t) + \xi(t) \right) \nonumber
\end{align}
    }
where we also account for the fact that, at the fixed point time correlations
are constant so that $G(t, t') = G(t - t', 0)$, which we rewrite as $G(t - t')$.

Since (\ref{eqn::perturbeddynamics}) contains a convolution term
$\int dt' G(t - t') \tilde{x}(t')$,
a classical eigenvalue analysis is intractable as an approach to determining stability. Instead, we adopt the
procedure presented in \cite{opper:phase} which we now outline, with full details
given in the Supplementary Material. We first take the
Fourier transform of (\ref{eqn::perturbeddynamics}). Doing so yields an
equation in terms of frequency $\omega$ rather than time $t$ and reduces the
convolution term into a product. In particular we obtain
\begin{align} 
    A(\omega, N_0) x(\omega) &= \sqrt{N_0} \eta(\omega) + \xi(\omega). \label{eqn::fourierdynamics} \\ 
    A(\omega, N_0) &= \left[ \frac{i \omega + T}{\bar{x}} - N_0 \Gamma G(\omega) \right] \nonumber
\end{align}
where we overload notation by identifying each variable with its Fourier
Transform, e.g. $\eta(\omega) = \mathcal{F}(\tilde{\eta})(t)$. This leads to the relation
\begin{align*}
    \langle |x(\omega)|^2 \rangle_* &= N_0 \Big(\langle |\eta(\omega)|^2| \rangle_* + 1\Big) \left \langle \frac{1}{|A(\omega, N_0)|}^2 \right\rangle_*
\end{align*}
where we recall again that $\langle \cdot \rangle_*$ denotes an expectation over all realisations of
the effective dynamics from an ensemble of games drawn with the same choice of $\Gamma$. In order to
analyse asymptotic stability, we focus on the limit $\omega \rightarrow 0$, since this corresponds
to longer timescales in $t$. Finally, we apply the relation $\langle \eta(t) \eta(t') \rangle_* =
\langle x(t) x (t') \rangle_*$ to write the dynamic solely in terms of $x$. This gives
\begin{align}
    \langle |x(\omega = 0)|^2 \rangle_* &= \Big( \frac{1}{\langle \frac{1}{| A(\omega=0, N_0) |}^2 \rangle_*} - N_0 \Big)^
     {-1} \label{eqn::lowfreq}
\end{align}
By definition, the left hand side of (\ref{eqn::lowfreq}) is positive, so a
contradiction is reached if 
\begin{equation}\label{eqn::stabilitycond} 
    N_0^{-1} < \left \langle \frac{1}{\left| \frac{T}{\bar{x}} - N_0 \Gamma \chi \right|^2} \right \rangle_*.
\end{equation}
where $\chi = \int_0^\infty G(s) ds$. As a result, (\ref{eqn::stabilitycond})
defines a sufficient condition for the onset of instability in the effective dynamics.

\subsection{Discussion}
\begin{figure}[t!]
    \centering
    \includegraphics[width=0.85\columnwidth]{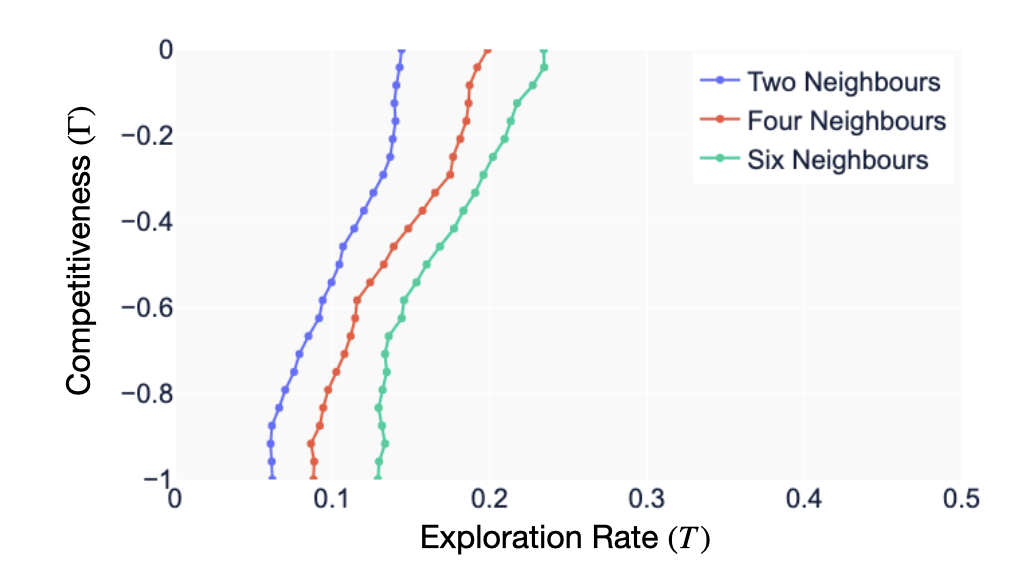}
    \caption{\label{fig::theory-plot} Stability boundary in terms of payoff correlation ($\Gamma$) and exploration
    rate ($T$) as given by the condition (\ref{eqn::stabilitycond}). Parameter choices on the right of the
    line satisfy the necessary condition for stability whilst those on the left fail.}
\end{figure}

In general, the stability condition (\ref{eqn::stabilitycond}) is not straightforward to parse and
cannot be solved in closed form. This is mostly due to the dependence on the
fixed point $\bar{x}$, whose form can be complicated. Nevertheless, it is
possible to numerically estimate the location of the stability boundary, i.e., the boundary between
choices of $(\Gamma, T)$ which satisfy (\ref{eqn::stabilitycond}) and those which do not. To do this, we
fix a choice of $\Gamma, T$, iteratively solve for $\bar{x}$ and subsequently evaluate
(\ref{eqn::stabilitycond}). Repeating this procedure for many choices of $\Gamma, T$ yields the stability
boundary depicted in Figure \ref{fig::theory-plot}, in which each line depicts the transition from satisfying 
the condition (\ref{eqn::stabilitycond}) on the right to its violation on the left. By examining this we can
assess how each of the parameters influence the stability of Q-Learning Dynamics.

The most notable feature of the necessary condition for stability
(\ref{eqn::stabilitycond}) is the dependence on the number of neighbours $N_0$. Even from
(\ref{eqn::stabilitycond}) itself we can discern the explicit independence on the total
number of agents $N$, which does not appear anywhere in the condition. Therefore, in competitive games,
the stability of learning is not influenced by the total number of agents, so long as the number of neighbours
per agent is kept constant. By contrast, as the number of neighbours increases the unstable region occupies 
more of the parameter space.

This result refines that of \cite{sanders:chaos} in which a stability boundary was determined for multi-player
games without any underlying communication structure. In this setting the authors showed that increasing
number of players leads to a larger unstable region. In our setting, this corresponds to a fully connected
network in which $N_0 = N - 1$. From Figure \ref{fig::theory-plot} it is clear that our result yields the
same prediction. Similarly, (\ref{eqn::stabilitycond}) aligns
exactly with the result of \cite{galla:complex} which derives a stability boundary in two-player games. In our
case this corresponds to any network game in which $N_0 = 1$.

Another feature shown by Figure \ref{fig::theory-plot} is that the stability
boundary increases in $T$ as $\Gamma$ decreases from $-1$ to $0$. That is, as the strength of anticorrelation between
agent payoffs decreases, higher exploration rates are required for Q-Learning dynamics to settle to an
equilbrium. Recall that $\Gamma =
-1$ corresponds to case where games along each edge are exactly negatively correlated, i.e.,
a zero-sum game. From \cite{piliouras:zerosum} it is known that
that Q-Learning dynamics asymptotically converge in network games which are exactly
zero-sum for any positive value of $T$. Figure \ref{fig::theory-plot} shows
that, as the competitiveness of the game decreases (i.e., as
$\Gamma \rightarrow 0$), higher exploration rates are required to guarantee convergence.

% Finally, we recall that, in order to derive the effective dynamics
% (\ref*{eqn::effectivedynamics}), we analysed the limit of large payoff matrices
% ($n \rightarrow \infty$). This is the setting in which the stability boundary
% (\ref*{eqn::stabilitycond}) holds exactly. Naturally, this overestimates the
% stability boundary in games of finite actions. However, our experiments show
% that the predictions made in the limit hold in practice for finite games.

\section{Experiments} \label{sec::Experiments}

\begin{figure*}[t!]
	\centering
	\begin{subfigure}{0.45\textwidth}
		\centering
		\includegraphics[width=0.85\columnwidth]{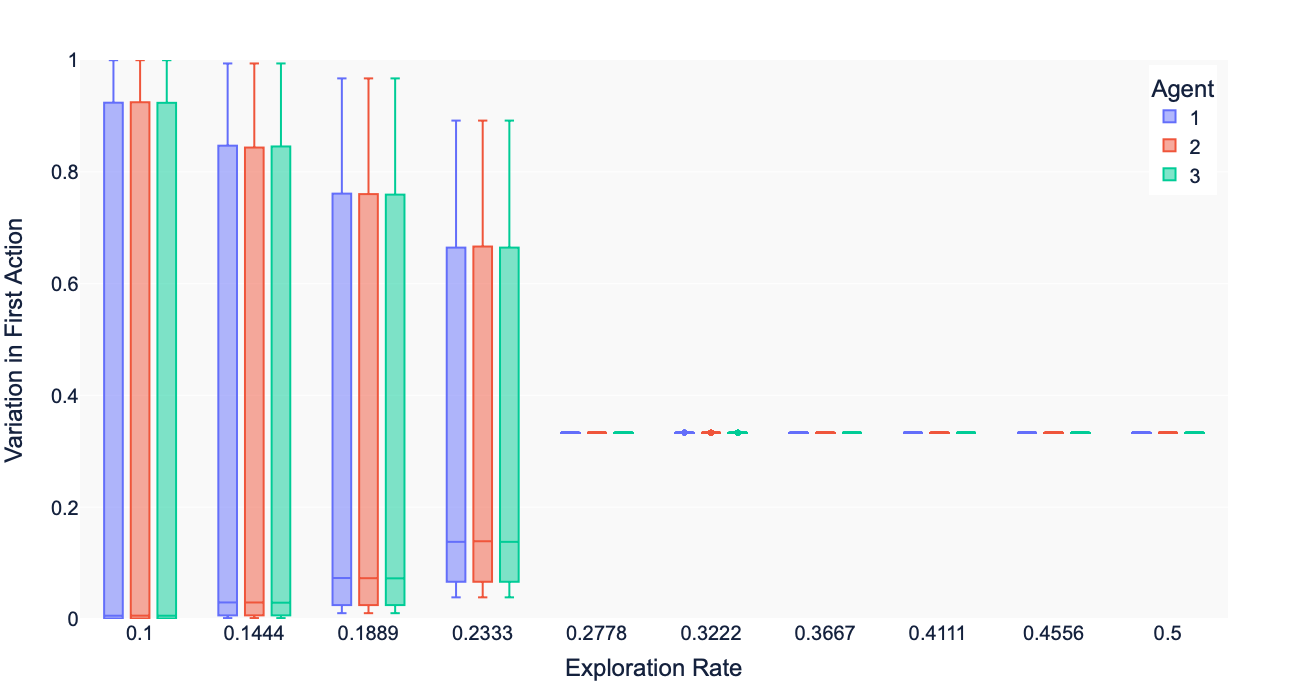}
		\caption{\label{fig::sato-boxplot-ring} Ring Network}
	\end{subfigure}
	\begin{subfigure}{0.45\textwidth}
		\centering
		\includegraphics[width=0.85\columnwidth]{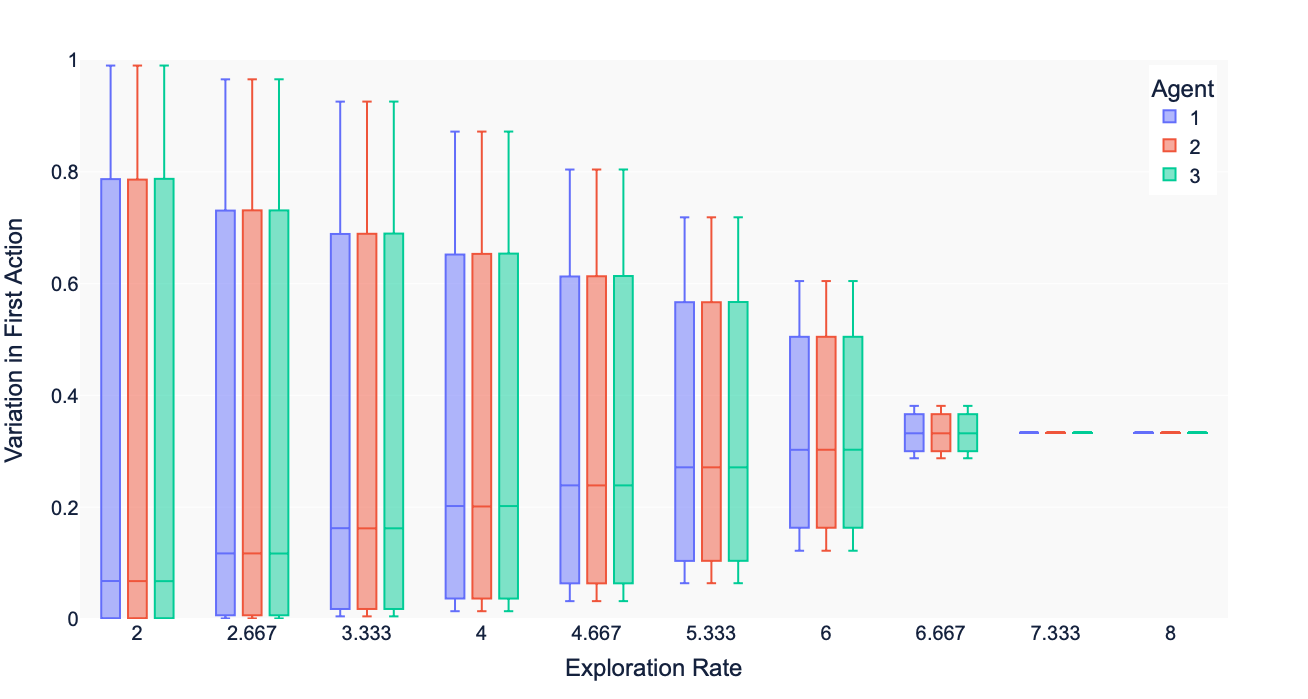}
		\caption{\label{fig::sato-boxplot-full} Full Network}
	\end{subfigure}        
	\caption{\label{fig::sato-boxplot} Boxplot depicting the final 2500
		iterations of learning in a Network Sato Game with 50 agents for various
		values of $T$. The mixed strategies of three agents are plotted depicting
		the spread of the trajectories across the simplex.}
\end{figure*}

\begin{figure*}[t!]
	\centering
	\begin{subfigure}{0.45\textwidth}
		\centering
		\includegraphics[width=\columnwidth]{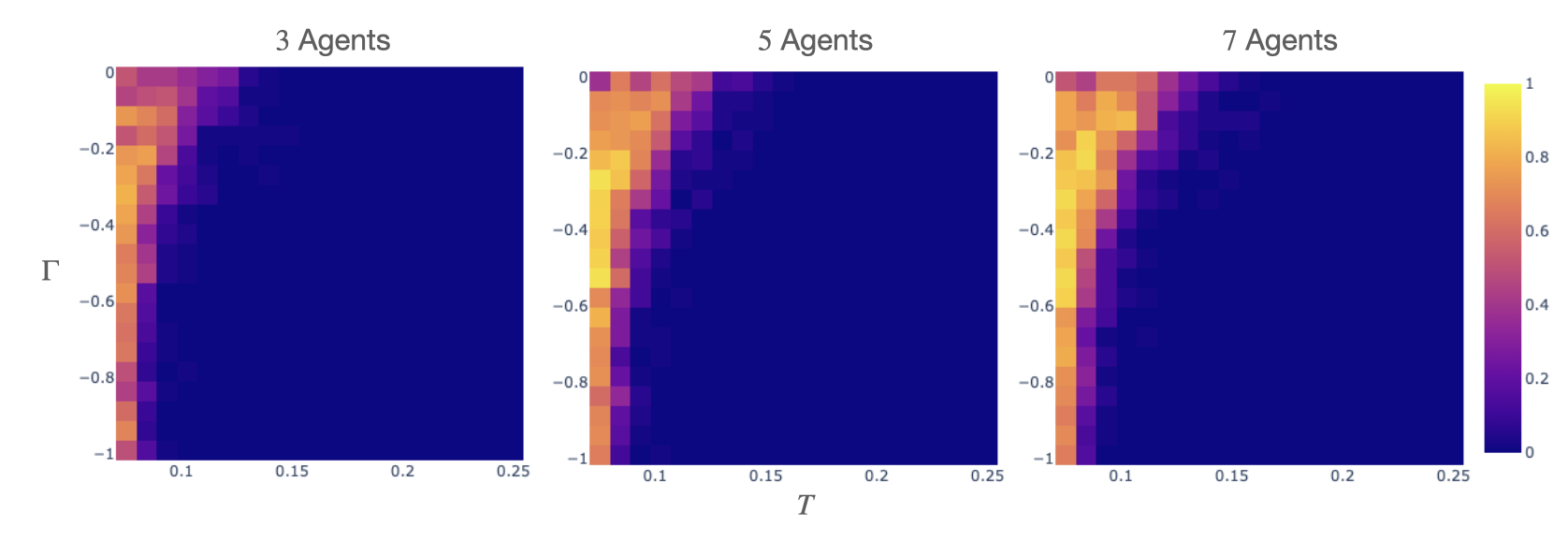}
		\caption{\label{fig::sato-traj-ring} Ring Network}
	\end{subfigure}
	\begin{subfigure}{0.45\textwidth}
		\centering
		\includegraphics[width=\columnwidth]{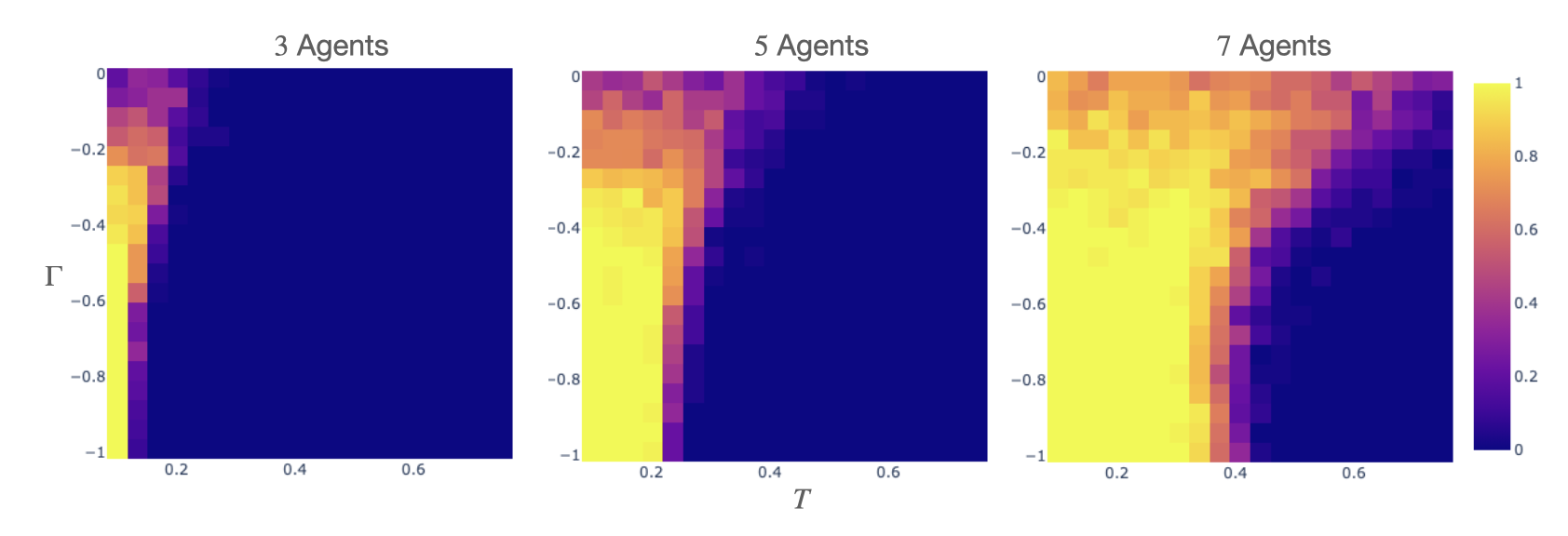}
		\caption{\label{fig::sato-traj-full} Full Network}
	\end{subfigure}        
	\caption{Empirically determined probability of non-convergence
		for various choices of $\Gamma, T$. Hot colours denote that a higher fraction of
		randomly drawn games fail to reach an equilibrium whilst cool colours
		depict a higher probability of convergence.} \label{fig::heatmaps}
\end{figure*}

% \begin{figure*}[t]
	%     \centering
	%     \begin{subfigure}[b]{0.3\textwidth}
		%         \centering
		%         \includegraphics[width=0.75\textwidth]{Experiments/Figures/Heatmap_Ring_3.png}
		%     \end{subfigure}
	%  %
	%     \begin{subfigure}[b]{0.3\textwidth}
		%         \centering
		%         \includegraphics[width=0.75\textwidth]{Experiments/Figures/Heatmap_Ring_5.png}
		%     \end{subfigure}
	%  %
	%  \begin{subfigure}[b]{0.3\textwidth}
		%     \centering
		%     \includegraphics[width=0.75\textwidth]{Experiments/Figures/Heatmap_Ring_7.png}
		% \end{subfigure}
	
	%  \caption*{Stability Boundary in the Ring Network}
	
	%     \begin{subfigure}[b]{0.3\textwidth}
		%         \centering
		%         \includegraphics[width=0.75\textwidth]{Experiments/Figures/Heatmap_Full_3.png}
		%     \end{subfigure}
	%  %
	%     \begin{subfigure}[b]{0.3\textwidth}
		%         \centering
		%         \includegraphics[width=0.75\textwidth]{Experiments/Figures/Heatmap_Full_5.png}
		%     \end{subfigure}
	%  %
	%  \begin{subfigure}[b]{0.3\textwidth}
		%     \centering
		%     \includegraphics[width=0.75\textwidth]{Experiments/Figures/Heatmap_Full_7.png}
		% \end{subfigure}
	
	%  \caption*{Stability Boundary in the Full Network}
	
	%  \caption{Empirically determined probability of non-convergence
		%  for various choices of $\Gamma, T$. Hot colours denote that a higher fraction of
		%  randomly drawn games fail to reach an equilibrium whilst cool colours
		%  depict a higher probability of convergence.} \label{fig::heatmaps}
	% \end{figure*}

\begin{figure*}[t!]
	\centering
	\begin{subfigure}{.45\textwidth}
		\includegraphics[width=\columnwidth]{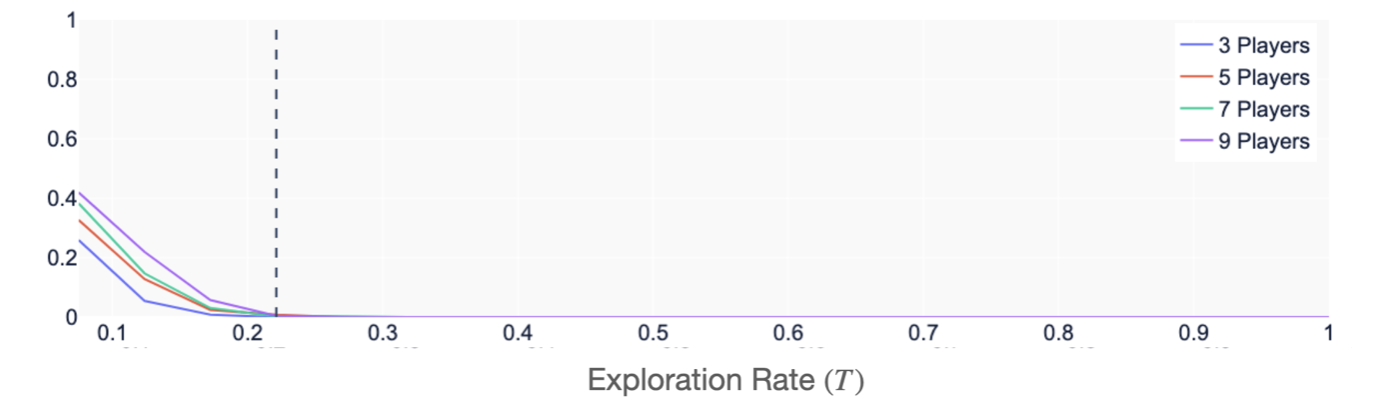}
	\end{subfigure}
	\begin{subfigure}{.45\textwidth}
		\includegraphics[width=\columnwidth]{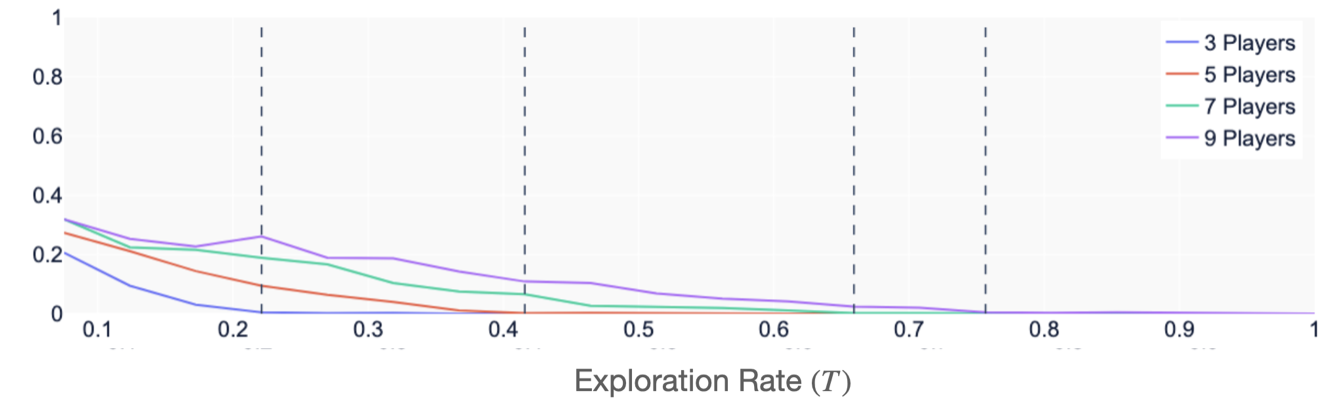}
	\end{subfigure}        
	\caption{\label{fig::stability-boundary} Probability of non-convergent
		dynamics with respect to $T$ over 250 randomly generated games with
		random choice of $\Gamma$. The black line depicts the choice of $T$ for
		which all games converge to a fixed point. In the ring network (Left), this
		occurs at a single value of $T$, whilst in the fully connected network (Right),
		the choice of $T$ depends on the number of players.}
\end{figure*}   

In this section we test the validity of the predictions made in our theoretical
analysis in the case of games with finite action sets by running the Q-Learning algorithm,
outlined in the Preliminaries.

% \begin{figure}[t!]
%     \centering
%     \begin{subfigure}{0.45\columnwidth}
%         \centering
%         \includegraphics[width=0.75\columnwidth]{Experiments/Figures/ring-graph.png}
%         \caption*{$N_0 = 2$}
%     \end{subfigure}        
%     \begin{subfigure}{0.45\columnwidth}
%         \centering
%         \includegraphics[width=0.75\columnwidth]{Experiments/Figures/full-graph.png}
%         \caption*{$N_0 = N - 1$}
%     \end{subfigure}
%     \caption{\label{fig::example-networks} Representative examples of regular networks studied in Section
%     \ref*{sec::Experiments} and associated $N_0$. (Left) Ring Network (Right) Full Network}
% \end{figure}

\paragraph{Representative Examples of Networks} In our experiments we analyse
two examples of networks - the \emph{ring network} and the \emph{fully connected
network}. These act as
prototypical examples for regular networks, which satisfy Assumption \ref{ass::regular-network}. In the
former, the payoff for any agent $k$ is given by
\begin{equation*}
    u_k(\x_k, \x_{-k}) = \x_k^\top A^{k, k-1} \x_{k-1} + \x_k^\top A^{k, k+1} \x_{k+1}
\end{equation*}
where addition and subtraction are taken $\mod N$. In this case, each agent has
only two neighbours, i.e. $N_0 = 2$ so the network connectivity is independent
of the total number of agents $N$. In the fully connected network, the payoff is
given by
\begin{equation*}
    u_k(\x_k, \x_{-k}) = \sum_{l \neq k} \x_k^\top A^{kl} \x_l
\end{equation*}
so that each agent has $N_0 = N-1$ neighbours. This corresponds also to the case
analysed by \cite{sanders:chaos} and \cite{hussain:aamas} in which it was
predicted that the boundary between stable and unstable learning dynamics is
impacted by the total number of agents.

% \paragraph{Example: Network Sato Game} We first illustrate the behaviour of
% Q-Learning on a representative example. This
% is an extension of the variant of Rock-Paper-Scissors first examined in
% \cite{sato:rps}. We extend this to the network
% setting by assuming that each edge $(k, l) \in \edgeset$ defines the same
% bimatrix game $(A, B)$, i.e. $(A^{kl}, A^{lk}) = (A, B)$ where
% \begin{align*} 
%     A=\begin{pmatrix} \epsX & -1 & 1 \\
%     1 & \epsX & -1 \\
%     -1 & 1 & \epsX \end{pmatrix}, \, B=\begin{pmatrix} \epsY & -1 & 1 \\
%     1 & \epsY & -1 \\
%     -1 & 1 & \epsY \end{pmatrix},
% \end{align*}   
% with $\epsX, \epsY \in \R$. Notice that if $\epsX=-\epsY$, the game is zero-sum.
%  In \cite{sato:rps,sato:qlearning} the case $\epsX =
% 0.1, \epsY = -0.05$ was analysed, and chaotic learning dynamics were found. In
% Figure \ref*{fig::sato-qlearning}, we plot trajectories generated by Q-Learning
% a seven agent network using a ring network and a fully connected network. We
% iterate Q-Learning for $50,000$ iterations and plot the mixed strategies of a
% single representative agent. In the ring network, it is clear
% that the dynamics do not converge for low choices of $T$. In particular, for $T
% = 0.1$, Q-Learning displays the same chaotic behaviour shown in \cite{sato:rps}
% for the two-player case. At $T = 0.35$ Q-Learning reaches the unique QRE at the
% uniform distribution $\NE_k = (1/3, 1/3, 1/3)$. By contrast the fully connected
% network does not converge for $T = 0.35$ but rather remains around the boundary
% of the simplex. Instead $T = 1$ is required to reach the QRE.

\paragraph{Example: Network Sato Game} We first illustrate the behaviour of
Q-Learning on a representative example. This
is an extension of the variant of Rock-Paper-Scissors first examined in
\cite{sato:rps}. The network extension is described in the Supplementary Material, alongside visualisations of
chaotic trajectories generated by Q-Learning.
In Figure \ref*{fig::sato-boxplot}, we
simulate 50 agents playing the Network Sato Game. We record
the agents' mixed strategies in the final $2500$ iterations of Q-Learning and,
for three representative agents, plot the probabilities with which they play
their first action. As such, Figure \ref*{fig::sato-boxplot} depicts the
spread of the asymptotic trajectory over the simplex. It is clear that, in the
fully connected network, a large value of $T$ is required in order for the
agents to converge to an equilibrium, whereas in the ring network $T \approx
0.3$ is sufficient. 

% Comparing this with Figure \ref*{fig::sato-qlearning}, the
% increase in the number of agents had no effect on convergence in the ring
% network, whereas in the fully connected network, the asymptotic behaviour remains
% non-stationary for comparatively large values of $T$.

\paragraph{Arbitrary Finite Games} Next, we determine the correctness of the analytic result in
arbitrary games. In Figure \ref*{fig::heatmaps}, we draw $50$ games with $50$ actions for each agent given choice of
$\Gamma$ using the formulation in \ref{eqn::payoffcorrelations}. Once again, we simulate Q-Learning
dynamics for $75,000$ time steps and record the final $10000$ iterations. To characterise the
limiting behaviour, we apply the following heuristic. We first determine whether, for each agent and
each strategy component, the relative difference between the maximum and minimum value across all
10000 time-steps is less than $0.01$. Formally, we determine whether 
\begin{equation*}
    \frac{\max_{t'} x_{ki}(t') - \min_{t'} x_{ki}(t')}{\max_{t'} x_{ki}(t')} < 0.01
\end{equation*}
where $t'$ is taken over the final 10000 iterations of learning.
Next, we determine the variance across the final iterations as
\begin{equation}\label{eqn::Variance} V = \frac{1}{Nn} \sum_{k,i} \frac{1}{10000} \sum_t x_{ki}(t)^2 - \left [\frac{1}{10000} \sum_t x_{ki}(t) \right]^2
\end{equation}

We check if the variance is less than $1 \times 10^{-5}$. If both of the above conditions are met,
then the dynamic is considered to have converged. As an example, the convergent
dynamics in Figure \ref*{fig::sato-boxplot} satisfy both of these conditions.
If, across all 10000 time steps, there is some $\tau$ such that, for all
agents $k$ and all $i \in
\actionset{k}$, $x_ {ki}(0)$, $x_{ki}( \tau)$, $x_{ki}(2 \tau)$, $x_{ki}(3 \tau)$ are all within
$0.01$ of each other, then it is considered that a stable periodic orbit has been reached. 

If neither of these conditions are met, then the dynamic is deemed to be non-convergent. Note this
does not conclude that the dynamics are formally chaotic, as limit cycle behaviours are also known
to occur in learning \cite{piliouras:cycles,imhof:cycles}. Examples of such behaviours are displayed in the Supplementary Material. However, in \cite{galla:complex,sanders:chaos} Q-Learning dynamics were
shown to exhibit chaotic behaviour for certain choices of $\Gamma, T$. This is also in line with the
rich literature on chaos in multi-agent learning \cite{sato:rps,chakraborty:chaos,svs:chaos}. 

From Figure \ref{fig::heatmaps}, it can be seen that the form of the analytic stability
boundary holds in practice. The reason
that the empirical boundary overestimates the theoretical result is that the latter considers
asymptotic trajectories, i.e. infinite time-scales, whilst the former is evaluated over 75,000
steps. Therefore, in the experiments, slow convergence can be mistaken for non-convergence. Overall, however, we see a strong alignment between the theoretical predictions and experimental evaluation. Namely, lower exploration rates are required strictly competitive games, i.e. as $\Gamma \rightarrow -1$, whilst uncorrelated games require higher exploration rates in order to converge to an equilibrium. In
addition, the stability boundary is unaffected by the number of players in the
ring network. 

The latter point is highlighted in Figure \ref{fig::stability-boundary} in
which we average over multiple choices of $\Gamma$ to isolate the effect of the
number of neighbours $N_0$ on the stability boundary in terms of $T$. It is
evident that increasing the number of players in a ring network plays no impact
on the stability boundary. As a result, it is possible to increase the number of
agents in such games arbitrarily without compromising convergence of Q-Learning
to a fixed point. By contrast fully connected networks do not scale well, as
non-convergent behaviours remain prevalent for low values of $T$ as the number
of agents increase.

\section{Conclusion}

In this study we have refined the previously held belief that chaotic behaviours
are more prevalent in games with many players. In particular, we analyse network
games and show that stability of the Q-Learning dynamic depends on the structure 
of the network, the competitiveness of the game and the exploration rates of agents. We show that in certain
networks, such as the ring network, an arbitrary number of agents may be added
to the system without compromising the propensity for learning to converge. By
contrast, if agents are heavily connected in the network, non-convergent
behaviours, such as limit cycles and chaos, become prevalent even with a small
total number of agents.

The present work has isolated the effect of the number of neighbours and the number of players in
the game. However, there are other factors which may affect the stability of Q-Learning. For
instance, whilst we have required $N_0$ to be the same for all agents, it would be fruitful to
analyse heterogeneously coupled networks. More generally, tools from graph
theory may be applied within this framework to uncover the role of the network
in convergence for arbitrary network games. Our work, therefore, presents a first step
towards building a complete picture of the stability of multi-agent learning in
network games.

\section*{Acknowledgments}
Aamal Hussain and Francesco Belardinelli are partly funded by the UKRI Centre for Doctoral Training in Safe and Trusted Artificial Intelligence (grant number EP/S023356/1)

% \section{Acknowledgments}
% AAAI is especially grateful to Peter Patel Schneider for his work in implementing the original aaai.sty file, liberally using the ideas of other style hackers, including Barbara Beeton. We also acknowledge with thanks the work of George Ferguson for his guide to using the style and BibTeX files --- which has been incorporated into this document --- and Hans Guesgen, who provided several timely modifications, as well as the many others who have, from time to time, sent in suggestions on improvements to the AAAI style. We are especially grateful to Francisco Cruz, Marc Pujol-Gonzalez, and Mico Loretan for the improvements to the Bib\TeX{} and \LaTeX{} files made in 2020.

% The preparation of the \LaTeX{} and Bib\TeX{} files that implement these instructions was supported by Schlumberger Palo Alto Research, AT\&T Bell Laboratories, Morgan Kaufmann Publishers, The Live Oak Press, LLC, and AAAI Press. Bibliography style changes were added by Sunil Issar. \verb+\+pubnote was added by J. Scott Penberthy. George Ferguson added support for printing the AAAI copyright slug. Additional changes to aaai24.sty and aaai24.bst have been made by Francisco Cruz and Marc Pujol-Gonzalez.

% \bibliographystyle{aaai24}
\bibliography{references}

\appendix

\onecolumn

In this Appendix we provide further details into the work presented in the main paper. In particular, we
present the full derivation of the stability boundary. To do this, we first outline our main tools which include:
the Q-Learning Dynamics (Section 2), the Generating Functional Approach (Section 3.1) and Fixed Point Analysis (Section 3.4).
% We also present additional experiments on our results (Section ), importantly predicting the existence of chaotic dynamics
% in games with low exploration rates.

\section{Q-Learning Dynamics}

In Q-Learning \cite{Watkins:qlearning,sutton:barto} an agent is required to determine an optimal strategy
through repeated interactions with its opponents. To do this, the agent keeps track of \emph{Q-values}, which 
are estimates of the reward associated with playing a given action. Any agent $k$ updates the Q-value of an action
$i$
\begin{equation}
	Q_{ki}(t + 1) = (1 - \alpha_k) Q_{ki}(t) + \alpha_k r_{ki}(\x_{-k})
\end{equation}
where $\alpha_k$ is a step size and $r_{ki}(\x_{-k})$ is the expected reward to $k$ for playing action $i$
when its opponents play the joint strategy $\x_{-k}$. Then, the agent updates their strategy
according to
\begin{equation}
	x_{ki}(t) = \frac{e^{Q_{ki}(t) / T_k}}{\sum_j e^{Q_{kj}(t) / T_k}}    
\end{equation}
where $T_k$ is the \emph{exploration rate} of agent $k$. In \cite{tuyls:qlearning} it is found that a variant of the popular \emph{replicator dynamics} acts as a model for the long term behaviour of Q-Learning. We call this dynamic the \emph{Q-Learning dynamics} and it is given as follows
\begin{equation} \tag{QLD} 
	\frac{\dot{x}_{k i}}{x_{k i}}=r_{k i}(\x_{-k})-\langle \mathbf{x}_k, r_k(\x_{-k}) \rangle +T \sum_{j \in S_k} x_{k j} \ln \frac{x_{k j}}{x_{k i}},
\end{equation}
For the purposes of this study, we rewrite (\ref{eqn::QLD}) in the following equivalent form
\begin{align}
	\frac{\dot{x}_{ki}}{x_{ki}}&=r_{ki}(\x_{-k}) - T \ln x_{kj} - \rho_k  \\
	\rho_k &= \langle \x_k, r_k(\x_{-k}) \rangle - T \langle \x_k, \ln \x_k \rangle \nonumber \\ 
	r_{ki}(\x_{-k}) &=  \sum_{l \in N_k} \left(A^{kl} \x_l \right)_i.
\end{align}
in which $\rho_k$ is a normalisation term which ensures that all strategy components $x_{ki}$ sum to $1$. We use this dynamic as the starting point for our analysis of Q-Learning.

\section{Generating Functional Approach}

In this section, we give a brief introduction to the use of Generating Functions as an approach to
analysing disordered systems. To this end, we follow the exposition presented in
\cite{coolen:tutorial} which provides further details for the interested reader.

\begin{example}
	Consider $N$ particles each with associated state $a_k$. The full state of the system is
	described as $a = (a_k)_{k = 1, \ldots, n}$. At any time step $t$, the system may occupy the
	state $a(t)$ with probability $x_t(a)$ which is to be updated as
	\begin{equation*}
		x_{t+1}(a) = \sum_{a'} W_t[a, a', \theta(t)] x_t(a'), \hspace{1 cm}
	\end{equation*}
	in which $W_t[a, a', \theta(t)]$ denotes the probability transition from a state $a$ to $a'$
	under the presence of an external, time varying force $\theta(t)$. We denote the probability of
	a given \emph{path} $(a(t))_{t = 0}^T$ as
	\begin{equation*}
		P[(a(t))_{t = 0}^T] = W_{T-1}[a, a', \theta(T-1)] \ldots W_0[a, a', \theta(0)]x_0(a(0))
	\end{equation*}
	With this we can define the \emph{generating functional} as 
	\begin{equation}
		Z(\varphi) = \sum_{a(0)} \ldots \sum_{a(t)} P[(a(t))_{t = 0}^T] e^{-i \sum_s \sum_k \varphi_k(s) a_k(s)}
	\end{equation}
	where $\varphi$ is an arbitrary source field. In order to be physically meaningful, the source
	field must be set to zero at the end. The generating functional describes the statistics of
	paths in that expectations, time correlations and response functions can be derived directly
	from $Z$.
	\begin{align*}
		\mathbb{E}[a_k(s)] &= i \lim_{\phi \rightarrow 0} \frac{\partial Z}{\partial \phi_k(s)}, \nonumber \\  
		\mathbb{E}[a_k(s) a_l(s')] &= - \lim_{\phi \rightarrow 0} \frac{\partial^2 Z}{\partial \phi_k(s) \phi_l(s')},  \nonumber \\
		\frac{\partial}{\partial \theta_l(s')}\mathbb{E}[a_k(s)] &= i \lim_{\phi \rightarrow 0} \frac{\partial^2 Z}{\partial \phi_k(s) \theta_l(s')}
	\end{align*}
\end{example}

The presented example motivates the use of the generating functional, which captures the statistics
of path probabilities. 

\section{Rescaling Of Variables}

In the case of the Q-Learning dynamics, the transition from $\x(t)$ and $\x(t + 1)$ are given by
(\ref{eqn::QLD}). These are parameterised by the payoff matrices $(A^{kl}, A^{lk})_{k, l} \in
\edgeset$. In order to perform the stability analysis, we require taking the thermodynamic limit of
infinitely large payoff matrices, i.e. $n \rightarrow \infty$. 
In order to do this, we must
recognise that the expected rewards $r_{ki}(\x_{-k}) = \sum_{l \in N_k} \sum_{j \in S_l} A^{kl}_{ij} x_{lj}$
involves taking the sum over $n$ elements. The terms $x_{lj}$ can be assumed to be of order $1/n$, since they must sum to $1$. As for the payoff matrices, we have enforced that they are drawn from a multivariate normal distribution, whose covariance is independent of $n$. Hence, they 
are of order $n^0$. By the central limit theorem, then, the sum $\sum_{l \in N_k} \sum_{j \in S_l} A^{kl}_{ij} x_{lj}$ is of order $n^{-1/2}$. This means that, as $n$  increases, the differences between
the expected rewards for each action become less appreciable, tending to zero in the limit. As we are analysing the system
in this limit, we must ensure that the distinctions between actions remain meaningful. To do this, we will make the change of
variables
\begin{align*}
	A^{kl} &= \sqrt{n} \tilde{A}^{kl}\\
	\x_k &= \tilde{\x}_k / n,
\end{align*}
in which  $\tilde{A}^{kl}$ is now of order $n^{-1/2}$ and $ \tilde{\x}_k$ is order $n^0$. Then, the sum  $\sum_{l \in N_k} \sum_{j \in S_l} \tilde{A}^{kl}_{ij} \tilde{x}_{lj}$ is $O(n^0)$. In the transformed system, as $n$
increases, the expected payoff does not. In doing this, we have now enforced that 
\begin{align}\label{eqn::payoffcorrelations_normalised}
	\mathbb{E}[A^{kl}_{ij}] &= 0, \hspace{0.5cm} \forall k \in \agentset, \forall i, j \in \actionset{k}  \nonumber\\
	\mathbb{E}[\left(A^{kl}_{ij}\right)^2] &= 1 / n, \hspace{0.5cm} \forall k \in \agentset, \forall i, j \in \actionset
	{k} \\
	\mathbb{E}[\left(A^{kl}_{ij}\right) \left(A^{lk}_{ji}\right)] &= \Gamma / n, \hspace{0.5cm} \forall l \in \agentset_k, \forall
	i \in \actionset{k}, j \in \actionset{l} \nonumber
\end{align}
In order to ensure the consistency of (\ref{eqn::QLD}), we write
\begin{align*}
	T = \tilde{T}  n^{-1/2}
\end{align*}
where $\tilde{T}$ does not scale with $n$. In taking all of these together, we find that running
(\ref{eqn::QLD}) with $A^{kl}, \x_k, T$ is equivalent to running $\tilde{A}^{kl}, \tilde{\x}_k, \tilde{T}_k$. This can be seen
through the Q-Learning update given by
\begin{align*}
	Q_{ki}(t + 1) &= (1 - \alpha) Q_{ki}(t) + \alpha r_{ki}(\x_{-k}) \\
	&= (1 - \alpha) Q_{ki}(t) + \sum_{l \in \agentset_k} \sum_{j \in \actionset{l}} A^{kl}_{ij} x_{lj} \\
	&= (1 - \alpha) Q_{ki}(t) + n^{-1/2} \sum_{l \in \agentset_k} \sum_{j \in \actionset{l}} \tilde{A}^{kl}_{ij} \tilde{x}_{lj} \\
	\implies \tilde{Q}_{ki}(t + 1) &= (1 - \alpha) \tilde{Q}_{ki}(t) + \sum_{l \in \agentset_k} \sum_{j \in \actionset{l}} \tilde{A}^{kl}_{ij} \tilde{x}_{lj} 
\end{align*}
where $Q_{ki} = \tilde{Q}_{ki} n^{-1/2}$. This gives us the Q-update in the transformed system. Finally, using that $T = n^{-1/2} \tilde{T}$ we get
$\tilde{Q}_{ki} / \tilde{T} = Q_{ki} / T$ so that the Q-Learning update is unchanged. 

Throughout the remainder of this supplementary material, we perform our analysis with the scaled variables $\tilde{A}^{kl}, \tilde{\x}_k, \tilde{T}$ but, in
order to avoid confusion, we report experimental results in the main paper in terms of the original, unscaled
system $A^{kl}, \x_k, T$. For the sake of notational convenience, we will now drop the tilde notation. We will also,
for the time being, specify the action of agent $k$ as $i_k$. 

\section{Path Integral Analysis of Q-Learning}

In this section, we apply the generating functional approach to the Q-Learning dynamics. In this
case, rather than a discrete time update, the probability of action selection $\x(t)$ is updated via
(\ref{eqn::QLD}). Then, the generating functional $Z$ is given by
\begin{align*}
	Z(\varphi) = \int D[\x, \hat{\x}] \exp \left( i \sum_{k} \sum_{i_k \in \actionset{k}} \int dt \left[ \hat{x}_{k, \i_k} (\frac{\dot{x}_{k, i_k}(t)}{x_{k, i_k}(t)} + T \, \ln x_{k, i_k}(t) + \rho_k (t) - h_{k, i_k} (t)) \right] \right) \\
	\times \exp \left(-i \sum_{k, i_k} \int dt \left [\hat{x}_{k, i_k} \left ( \sum_{l \in N_k} A^{kl}_{i_k, i_l} x_{l, i_l} \right ) \right] \right) \\
	\times \exp \left(i \sum_{k, i_k}
	\int dt[x_{k, i_k} \varphi_{k, i_k} (t)] \right)
\end{align*}
where the $\hat{x}_{k, i_k}$ indicates the Fourier transform of $x_{k, i_k}$ and $h_{k, i_k} (t)$ (resp. $\varphi$) denotes an
external force (resp. source field) - the corollary of $\theta_k(t)$ (resp. $\varphi$) in our example - which too shall be set to zero. See
\cite{galla:complex} for similar calculations.

Our goal is to determine the \emph{effective dynamics}, which describes the Q-Learning dynamics
averaged over all payoff realisations from a choice of $\Gamma$. Our approach will be to determine
the averaged form of $Z$, which we call $Z_{\text{eff}}$ and then to identify a continuous time
dynamic which generates $Z_{\text{eff}}$. To do this, we first isolate the terms in $Z$ which
contain the random variables $A^{kl}_{i_k, i_l}$. We define
\begin{align*}
	\Pi =  \prod_{i_k : k \in \agentset} \, \exp(-i \sum_{k} \sum_{l \in \agentset_k} \int dt \, \hat{x}_{k, i_k}  \, A^{kl}_{i_k, i_l} \, x_{l, i_l})
\end{align*}
\subsubsection{Expectation of $\Pi$}

We first write $\Pi$ as 
\begin{equation*}
	\Pi =  \prod_{i_k : k \in \agentset} \exp(\mathbf{b} \cdot \mathbf{z}),
\end{equation*}
where
\begin{equation*}
	%    \begin{split}
		\mathbf{b} := [\ldots, -i \int dt \, \hat{x}_{k, i_k} x_{l, i_l},  \ldots]^T 
		%    \end{split}
\end{equation*}
is a vector which contains all of the permutations of the products $\hat{x}_{k, i_k} x_{l, i_l}$ and

\begin{equation*}
	%    \begin{split}
		\mathbf{z} := [ \ldots, A^{kl}_{i_k, i_l}, \ldots ]^T, 
		%    \end{split}
\end{equation*}
is a vector containing all of the payoff elements corresponding to the products in $\mathbf{b}$. Clearly, $\mathbf{z}$ is the random variable whose average we wish to determine. To do this, we apply following the exponential identity \cite{zinn:qft}
\begin{equation}
	\label{eqn::expectationIdentity}
	\int d \mathbf{z} [e^{-M_2(\mathbf{z}) + \mathbf{b} \cdot \mathbf{z}}] = \det(2 \pi M)^{-1/2} e^{\omega(\mathbf{b})},
\end{equation}
in which $M = \Sigma^{-1}$ and $\z \sim \mathcal{N}(\zeros, \Sigma)$, and
\begin{align*}
	%	\begin{split}
		M_2(\mathbf{z}) & = 1/2 \sum_{ij} z_i M_{ij} z_j \\
		\omega(\mathbf{b}) &  = 1/2 \sum_{ij} b_i (M)^{-1}_{ij} b_j\\
		M  & = \Sigma^{-1} \\
		%    \end{split}
\end{align*}
We recall that the scaled system has payoffs chosen so that
\begin{eqnarray}
	%    \begin{split}
		\mathbb{E} \left[ A^{kl}_{i_k, i_l} A^{lk}_{ji} \right] = \begin{cases} \frac{1}{n} &  \text{ if }
			l = k \\
			\frac{\Gamma}{n} & \text{ otherwise. }
		\end{cases}
		%    \end{split}
\end{eqnarray}
Applying the identity (\ref{eqn::expectationIdentity}) to $\Pi$ gives
\begin{eqnarray*}
	%\begin{split}
	\mathbb{E}[\Pi] = \exp \Big( \sum_{k, l \in \agentset_k} \sum_{i_k, i_l} - \frac{1}{2n} \int
	dt dt' & \hat{x}_{k, i_k}(t) x_{l, i_l}(t) \hat{x}_{k, i_k}(t') x_{l, i_l}(t') + x_{k, i_k}(t) \hat{x}_{l, i_l}(t)
	x_{k, i_k}(t') \hat{x}_{l, i_l}(t') \\
	+ & \Gamma \big[ \hat{x}_{k, i_k}(t) x_{l, i_l}(t) \hat{x}_{l, i_l}(t') x_{k, i_k}(t') + x_{k, i_k}(t)
	\hat{x}_{l, i_l}(t) \hat{x}_{k, i_k}(t') x_{l, i_l}(t') \big] \Big) 
	%\end{split}
\end{eqnarray*}
Next, we define the correlation functions
\begin{align*}
	%    \begin{split}
		C_k (t, t') & := n^{-1} \sum_{i} x_{k, i_k}(t) x_{k, i_k}(t') \\
		L_k (t, t') & := n^{-1} \sum_{i} \hat{x}_{k, i_k}(t) \hat{x}_{k, i_k}(t') \\
		K_k (t, t') & := n^{-1} \sum_{i} x_{k, i_k}(t) \hat{x}_{k, i_k}(t')
		%    \end{split}
\end{align*}
which allows us to rewrite the expectation as
\begin{equation}
	\begin{split}
		\mathbb{E}[\Pi] = \exp \left(- \frac{n}{2} \sum_{k} \sum_{l \in \agentset_k} \Big [ \int dt dt' L_k(t, t') C_l (t, t') +  \Gamma K_k (t, t') K_l (t', t) \Big ] \right)
	\end{split}
\end{equation}
To introduce these correlation functions into the integral, we rely on the use of their Dirac delta
functions in its Fourier transform, for example
\begin{equation}
	1 = \int D[C_k (t, t') \hat{C}_k (t, t')] \; \exp \left(in \int dt dt' \text{ } \hat{C}_k (t, t') (C_k (t, t') - n^{-1} \sum_{i_k} x_{k, i_k}(t) x_{k, i_k}(t')) \right)
\end{equation}
\subsection{The Effective Dynamics}
Having performed the expectation of the variables $A^{kl}$ in the generating functional, we can now
substitute these back in to yield
\begin{equation}
	\label{eqn::averagedgf}
	\mathbb{E}[Z(\varphi)] = \int D[C, \hat{C}, L, \hat{L}, K, \hat{K}] \text{ } \exp(n (P + \Phi + \Omega + \mathcal{O}(n^{-1}))),
\end{equation}
where
{\small
	\begin{align*}
		P &= i \sum_k \int dt dt' \left[ C_k (t, t') \hat{C}_k (t, t') + L_k (t, t') \hat{L}_k (t, t') + K_k (t, t') \hat{K}_k (t, t')]\right] \\
		\Phi &= - \frac{1}{2}  \sum_{k} \int dt dt' \left [  L_k(t, t') C_l (t, t') + \Gamma K_k (t, t') K_l (t', t) \right ] \\
		\Omega &= n^{-1} \sum_{k, i_k} \ln \Big \{ \int D[\x_k, \hat{\x}_k] p_{k, i_k}(0) \exp \left(i \int dt [x_{k, i_k}(t) \varphi_{k, i_k}(t)] \right) \\ &\times \exp \left(i\int dt \left[ \frac{\dot{x}_{k, i_k}(t)}{x_{k, i_k}(t)} + T \, \ln x_{k, i_k}(t) + \rho_k (t) - h_{k, i_k} (t)) \right] \right)  \\ &\times \exp \left(-i \int dt dt' \left[\hat{C}_k (t, t') x_{k, i_k} x_{k, i_k}(t') + \hat{L}_k (t, t') \hat{x}_{k, i_k}(t) \hat{x}_{k, i_k}(t') + \hat{K}_k (t, t') ) x_{k, i_k}(t) \hat{x}_{k, i_k}(t') \right]  \right)  \Big \}.
	\end{align*}
}

Here, $p_{k}(0)$ denotes the initial distribution which the initial mixed strategy of agent $k$ is
drawn from. It is important to note here that all of the information regarding the dynamics of the
agent is included within $\Omega$, in the sense that the integral over the strategy components $x$
are contained within this expression. 

We now wish to reduce the integral (\ref{eqn::averagedgf}) through the use of the saddle point
method of integration. In this method, we take the limit $n \rightarrow \infty$ so that the area
under the curve of (\ref{eqn::averagedgf}) is dominated by the maxima of the term $f = \Psi + \Phi +
\Omega $. Therefore, we first find the maxima of this term with respect to the
integral variables $[C, \hat{C}, L, \hat{L}, K, \hat{K}, A, \hat{A}]$. This yields {\small
	\begin{eqnarray*}
		\frac{\partial f}{\partial \hat{C}_k} = 0  & \implies & C_k (t, t') =  - \lim_{n \rightarrow \infty}
		n^{-1} \sum_i \mathbb{E}[x_{k, i_k}(t) x_{k, i_k}(t')]_{\Omega}  = - \lim_{n \rightarrow \infty} n^{-1}
		\sum_i \frac{\partial^2 \mathbb{E}[Z(\varphi)]}{\delta \varphi_{k, i_k}(t) \varphi_{k, i_k}(t')} \Big \rvert_{\varphi =
			\mathbf{h} = 0} \\
		\frac{\partial f}{\partial \hat{L}_k} = 0  & \implies & L_k (t, t') =  - \lim_{n \rightarrow \infty}
		n^{-1} \sum_i \mathbb{E}[\hat{x}_{k, i_k}(t) \hat{x}_{k, i_k}(t')]_{\Omega}  = - \lim_{n \rightarrow \infty}
		n^{-1} \sum_i \frac{\partial^2 \mathbb{E}[Z(\varphi)]}{\delta h_{k, i_k} (t) h_{k, i_k} (t')} \Big \rvert_{\varphi =
			\mathbf{h} = 0} \\
		\frac{\partial f}{\partial \hat{K}_k} = 0  & \implies & K_k (t, t') =  - \lim_{n \rightarrow \infty}
		n^{-1} \sum_i \mathbb{E}[x_{k, i_k}(t) \hat{x}_{k, i_k}(t')]_{\Omega} = - \lim_{n \rightarrow \infty} n^{-1}
		\sum_i \frac{\partial^2 \mathbb{E}[Z(\varphi)]}{\delta \varphi_{k, i_k} (t) h_{k, i_k} (t')} \Big \rvert_{\varphi =
			\mathbf{h} = 0},
	\end{eqnarray*}
}
where $\mathbb{E}[ \cdot ]_\Omega$ denotes an expectation to be taken over $\Omega$. The interested
reader may consult \cite{coolen:minority} for details on how such an expectation may be formulated.
However, for our purposes, the details are not required. What it is important to note, however, is
that by normalisation of the generating functional it was required that $Z(\varphi = 0, \mathbf{h}) = 1$ for any
choice of $\mathbf{h}$. Therefore, the $Z(\varphi)$ is constant in $\mathbf{h}$ and so $L_k (t, t') =
0$ for any $t, t'$. In addition, we know that past values of $x_{k, i_k}$ cannot be affected by future
values and so we have that $K_k(t, t') = 0$ for $t' > t$ and so $K_k(t, t') K_k(t', t) = 0$.

Continuing the extrema analysis we have
\begin{eqnarray*}
	\frac{\partial f}{\partial C_k(t, t')} = 0  & \implies & i \hat{C}_k (t, t') =  \frac{1}{2} \sum_{l \in
		\agentset_k} L_l (t, t') = 0 \\
	\frac{\partial f}{\partial L_k(t, t')} = 0  & \implies & i \hat{L}_k (t, t') = \frac{1}{2} \sum_{l \in
		\agentset_k} C_l (t, t')\\
	\frac{\partial f}{\partial K_k(t, t')} = 0 & \implies & i \hat{K}_k (t, t') = \, \Gamma \sum_{l \in
		N_k} K_l (t', t) 
\end{eqnarray*}
where the first equality holds since in $\Omega$, $C_k (t, t')$ always appears in a product with
$L_k (t, t')$ or $K_k(t, t') K_k(t', t)$, which we have already established to be 0. By substituting
these expressions back into (\ref{eqn::averagedgf}), we immediately get that $P + \Phi = 0$ since
each of the terms will contain one of the above terms which we found to be 0. For $\Omega$, we get
{\small
	\begin{align*}
		\Omega &= n^{-1} \sum_{k, i_k} \ln \Big \{ \int D[\x_k, \hat{\x}_k] p_{k, i_k}(0) \exp \left(i\int dt \left[ \frac{\dot{x}_{k, i_k}(t)}{x_{k, i_k}(t)} + T \, \ln x_{k, i_k}(t) + \rho_k (t) - h_{k, i_k} (t)) \right] \right)  \\ &\times \exp \left(-N_0 \int dt dt' \left[ \frac{1}{2} \hat{L}_k (t, t') \hat{x}_{k, i_k}(t) \hat{x}_{k, i_k}(t') + \Gamma \hat{K}_k (t, t')  x_{k, i_k}(t) \hat{x}_{k, i_k}(t') \right] \right)  \\ & \times \exp \left(i \int dt [x_{k, i_k}(t) \psi_{k, i_k}(t)] \right)  \Big \}
	\end{align*}
}

We make the assumption that all strategy components for all agents are drawn from the same initial
distribution $p_0$ and are under the influence of the same fields $h(t)$, $\psi(t)$. In doing so, 
we are able to drop the distinction between agents and their
action probabilities. Taking each degree of freedom of $\Omega$ to be an \emph{effective} generating
functional $Z_{\text{eff}}$, we get, after dropping the distinction between players and actions

\begin{eqnarray}
	\label{eqn::zeff}
	%\begin{split}
	Z_{\text{eff}} =& \int D[x, \hat{x}] p_{0} \exp \left(i\int dt \left[ \hat{x}(t)
	(\frac{\dot{x}(t)}{x(t)} + T \ln x(t) + \rho(t) - h(t)) \right] \right) \nonumber
	\\ &\times \exp \left(- N_0 \int dt dt' \left[ \frac{1}{2}  C(t, t')\hat{x}(t) \hat{x}(t') + i\Gamma
	G(t, t') x(t) \hat{x}(t') \right] \right) \\ & \times \exp \left(i \int dt [x(t) \psi(t)] \right) 
	%\end{split}
\end{eqnarray}
in which $G(t, t') = -iK(t, t')$. We notice that (\ref{eqn::zeff}) is exactly the form of the
generating functional which would generate the dynamics
\begin{equation}\label{eqn::effectivedynamics_sapp}
	\begin{split}
		\frac{1}{x(t)} \frac{d}{dt} x(t) = N_0 \Gamma \int dt' G(t, t') x(t')  - T \ln x(t) - \rho(t) + \sqrt{N_0} \eta(t), 
	\end{split}
\end{equation}
where
\begin{eqnarray*}
	%    \begin{split}
		\mathbb{E}[\eta(t)]_* & = & 1, \text{ \space } \mathbb{E}[\eta(t) \eta(t')]_*  =
		\mathbb{E}[x(t) x(t')]_* \\
		G(t, t') & = & \mathbb{E}\left [ \frac{\delta x(t)}{ \delta \eta(t')} \right]_*.
		%    \end{split}
\end{eqnarray*}

and we have also set $h = 0$. We therefore call (\ref{eqn::effectivedynamics_sapp}) the
\textit{effective dynamics}. This gives an approximate expression for the expected behaviour of a
strategy component averaged over all possible realisations of the payoff matrix. It is on this
system that we will perform a stability analysis.

\subsection{Fixed Points of the System}

Before we can analyse the stability of a fixed point, we first need to establish that the fixed
point exists for the system. At such a point $\dot{x}(t) = 0$. Furthermore, since there is no
fluctuation of $x(t)$ with respect to time, the term $C(t, t')$ is constant and $G(t, t') \approx
G(t - t')$, i.e. it becomes a function of time difference rather than being concerned of the exact
times $t, t'$. As $\eta$ is also both stationary at the fixed point, we rewrite $\eta$ as $q^{1/2}z$
where $q = \mathbb{E}[\bar{x}^2]$ and $z$ is drawn from the standard normal distribution (mean 0,
variance 1). With all of the above considered, (\ref{eqn::effectivedynamics_sapp}) reduces to the
fixed point equation
\begin{equation} \label{eqn::fixedpoints}
	0 = \bar{x} \Big[ N_0 \, \Gamma \bar{x} \chi - T \ln \bar{x} - \rho + \sqrt{N_0 q} z \Big]
\end{equation}
,where $\chi = \int G(t-t') dt $. Now, we have two possible choices for the fixed point: either
$\bar{x}$ is zero or $\bar{x}$ is given by the expression in the squared brackets. Since the QRE
are interior, we cannot have $\bar{x} = 0$. The fixed point
$\bar{x}$, then, is determined by solving the second term of (\ref{eqn::fixedpoints}). As decribed in \cite{opper:phase,galla:complex}, this
has a unique positive solution only when $\Gamma < 0$. Therefore, we must restrict our analysis to $\Gamma \in [-1, 0]$.

Of course, the expression in the brackets of (\ref{eqn::fixedpoints}) is not trivial to solve and so
a root finding method was employed to approximate the value. In doing so, one can approximate the
location of the fixed point $\bar{x}$. The parameters $q, \chi, \rho$ are determined as
\begin{eqnarray*}
	1 & = & \frac{1}{\sqrt{2 \pi}} \int_{-\infty}^{\infty} \bar{x}(z) \exp(-\frac{z^2}{2}) dz \\ 
	q & = & \frac{1}{\sqrt{2 \pi}} \int_{-\infty}^{\infty} \bar{x}(z)^2 \exp(-\frac{z^2}{2}) dz
	\\ 
	\chi & = & \frac{1}{q^{1/2}}\frac{1}{\sqrt{2 \pi}} \int_{-\infty}^{\infty} \frac{\partial
		\bar{x} (z)}{\partial z} \exp(-\frac{z^2}{2}) dz
\end{eqnarray*}

\subsection{Stability of the Effective Dynamics}

With the region $\Gamma < 0$ established and the fixed point examined, we now look at the behaviour
of the effective dynamics in a neighbourhood of the fixed point. In particular, we would like to
establish whether, if perturbed from $\bar{x}$, the system will diverge away (unstable) or converge
to $\bar{x}$ (stable). A similar question was considered in
\cite{opper:phase} and we progress along the same lines. 

The dynamics are proposed to be perturbed by a disturbance $\xi(t)$ which is drawn from a Gaussian
of zero mean and unit variance. The disturbance causes the values of $x(t)$ and $\eta(t)$ to deviate
from their fixed point position $\bar{x}$, $\bar{\eta}$ by an amount $\tilde{x}(t)$,
$\tilde{\eta}(t)$. Notice we have used a slight abuse in notation by using tildes. This is not to be
confused with the rescaling which took place earlier. Rewriting the dynamics with all of these
considerations yields
{\small
	\begin{eqnarray*}
		%   \begin{split}
			\frac{d}{dt} (\bar{x} + \tilde{x}(t)) &= (\bar{x} + \tilde{x}(t)) \Big (N_0 \Gamma \int
			dt' \left [G(t, t') (\bar{x} + \tilde{x}(t')) \right ] - T \ln (\bar{x} +
			\tilde{x}(t)) + \sqrt{N_0} (\bar{\eta} + \tilde{\eta}(t)) - \rho(t) + \xi(t) \Big )
			%  \end{split}
	\end{eqnarray*}
} In a neighbourhood of $\bar{x}$, the linear terms would dominate and so we keep these, neglecting
higher order terms. This yields {\small%
	\begin{eqnarray} \label{eqn::linearised}
		%    \begin{split}
			\frac{d}{dt} \tilde{x}(t) & = - T \tilde{x} (t) + (\bar{x} + \tilde{x}(t)) [ N_0 \Gamma \bar{x} \int
			dt' [ G(t, t') ] - T \ln \tilde{x}(t) - \rho + \sqrt{N_0} \bar{\eta}] \nonumber \\
			& + \bar{x} [N_0 \Gamma \int dt' [ G(t, t') \tilde{x}(t') ] + \sqrt{N_0} \tilde{\eta}(t) + \xi(t)]
			%    \end{split}
	\end{eqnarray}
} Now invoking the fixed point condition, we notice that the first square bracket in
(\ref{eqn::linearised}) equates to zero. Now, taking the Fourier transform of
(\ref{eqn::linearised}) gives
\begin{equation}
	\Big[ \frac{i \omega + T}{\bar{x}} - N_0 \Gamma \tilde{G}(\omega) \Big] x(\omega) + \sqrt{N_0} \eta(\omega) +  \xi(\omega)
\end{equation}
Using the relation  $\mathbb{E}[|\tilde{\eta}|^2]_* = \mathbb{E}[|\tilde{x}|^2]_*$, we have
\begin{eqnarray}
	\mathbb{E}[|x(\omega)|^2] & = \mathbb{E}[|\sqrt{N_0} \eta(\omega) +  \Xi(\omega)|^2]_*
	\mathbb{E} \left[ |A(\omega, N_0)|^{-2} \right]_* \nonumber \\
	& = \left( \frac{1}{\mathbb{E} \left[ |A(\omega, N_0)|^{-2} \right]_*} - N_0 \right)^{-1}
\end{eqnarray}
where
\begin{equation}
	A(\omega, N_0) = \frac{i \omega + T}{\bar{x}} - N_0 \Gamma G(\omega)
\end{equation}
As we wish to examine the long time behaviour of the system, we must consider the low frequency
$\omega = 0$ in order to remove all transients. Of course, $\mathbb{E}[|\tilde{x}(\omega=0)|^2]_* >
0$. Therefore, if we find, for a given choice of parameters, that
$\mathbb{E}[|\tilde{x}(\omega=0)|^2]_*$ yields a negative value, then the necessary condition for
stability is violated and we have the onset of instability. In particular, stability is violated if 

\begin{equation}
	\label{eqn::Final}
	\frac{1}{N_0} < \mathbb{E} \left|A(\omega=0, N_0)|^{-2} \right]_* =  \mathbb{E} \left[ |\frac{T}{\bar{x}} - N_0 \Gamma \chi|^{-2} \right]_*
\end{equation}

Finally, (\ref{eqn::Final}) is the expression whose predictions are analysed in the main body of the
paper. 

\section{Network Sato Game}

\begin{figure*}[t!]
	\centering
	\begin{subfigure}{0.45\textwidth}
		\centering
		\includegraphics[width=\columnwidth]{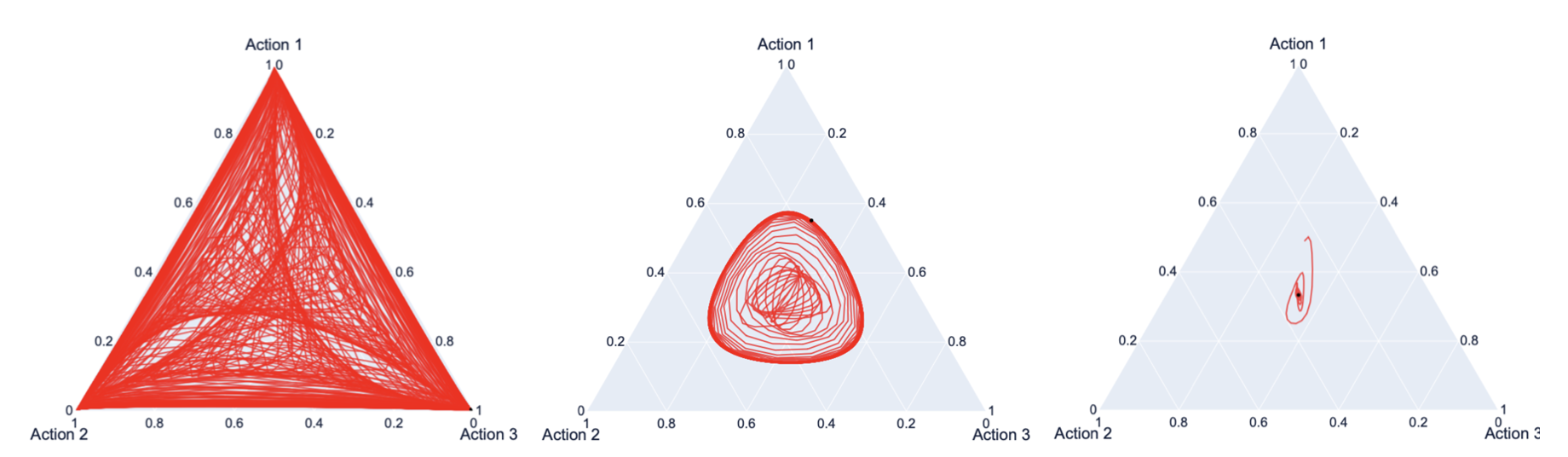}
		\caption{\label{fig::sato-traj-ring} Ring Network}
	\end{subfigure}
	\begin{subfigure}{0.45\textwidth}
		\centering
		\includegraphics[width=\columnwidth]{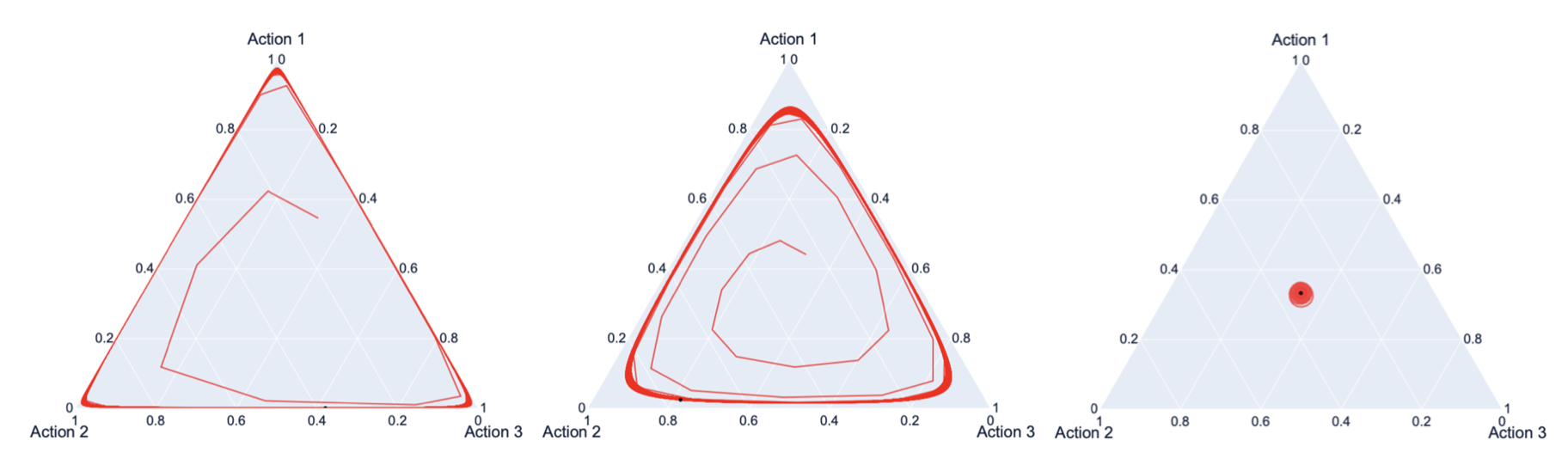}
		\caption{\label{fig::sato-traj-full} Full Network}
	\end{subfigure}        
	\caption{Trajectories of Q-Learning in Network Sato Game with seven
		agents. We plot the mixed strategies of the first agent as Q-Learning is
		applied for various choices of exploration rate $T$. In the ring network (Top)
		convergence to the QRE is achieved with $T = 0.35$, whilst in the fully
		connected network (Bottom) $T=1$ is required.} \label{fig::sato-qlearning}
\end{figure*}

In the main paper, we consider the \emph{Network Sato Game}, a multiplayer extension of the bimatrix game analysed in \cite{sato:rps}. In the network variant, each edge $(k, l) \in \edgeset$ defines the same
bimatrix game $(A, B)$, i.e. $(A^{kl}, A^{lk}) = (A, B)$ where
\begin{align*} 
	A=\begin{pmatrix} \epsX & -1 & 1 \\
		1 & \epsX & -1 \\
		-1 & 1 & \epsX \end{pmatrix}, \, B=\begin{pmatrix} \epsY & -1 & 1 \\
		1 & \epsY & -1 \\
		-1 & 1 & \epsY \end{pmatrix},
\end{align*}   
with $\epsX, \epsY \in \R$. Notice that if $\epsX=-\epsY$, the game is zero-sum.
In \cite{sato:rps,sato:qlearning} the case $\epsX =
0.1, \epsY = -0.05$ was analysed, and chaotic learning dynamics were found. In
Figure \ref*{fig::sato-qlearning}, we plot trajectories generated by Q-Learning
a seven agent network using a ring network and a fully connected network. We
iterate Q-Learning for $50,000$ iterations and plot the mixed strategies of a
single representative agent. In the ring network, it is clear
that the dynamics do not converge for low choices of $T$. In particular, for $T
= 0.1$, Q-Learning displays the same chaotic behaviour shown in \cite{sato:rps}
for the two-player case. At $T = 0.35$ Q-Learning reaches the unique QRE at the
uniform distribution $\NE_k = (1/3, 1/3, 1/3)$. By contrast the fully connected
network does not converge for $T = 0.35$ but rather remains around the boundary
of the simplex. Instead $T = 1$ is required to reach the QRE.

\end{document}